\documentclass[12pt]{article}
\usepackage{makeidx}
\usepackage{amsmath}
\usepackage{amsfonts}
\usepackage{amssymb}

\usepackage[utf8]{inputenc} 
\usepackage[T1]{fontenc}
\usepackage{textcomp} 
\usepackage[english]{babel}
\usepackage[autolanguage]{numprint} 
\usepackage{hyperref} 
\usepackage{graphicx} 
\usepackage{verbatim} 
\usepackage[all]{xy}
\usepackage{dsfont}
\usepackage{url}
\usepackage{graphicx}
\usepackage{amsmath,amssymb,amsthm} 
\usepackage{lmodern}
\usepackage{ tipa }
\usepackage{latexsym}
\usepackage{mathtools}
\usepackage{relsize}
\usepackage{exscale}
\usepackage[nottoc, notlof, notlot]{tocbibind}

\newtheorem{proposition}{Proposition}
\newtheorem{corollary}{Corollary}
\newtheorem{remark}{Remark}
\newtheorem{lemma}{Lemma}

\begin{document}

\title{ACCUMULATION OF INDIVIDUAL FITNESS OR WEALTH AS A POPULATION GAME}
\author{Sylvain Gibaud \\
%EndAName
Institut de Math\'{e}matique de Toulouse\\
Universit\'{e} Paul Sabatier, Toulouse\\
\medskip \and J\"{o}rgen W. Weibull\thanks{%
This manuscript builds upon an incomplete sketch by one of the authors, see
Weibull (1999).\ We thank Ingela Alger, J\'{e}r\^{o}me Renault, Laurent
Miclo, Daniel Waldenstr\"{o}m, and two anonymous referees for comments and
suggestions\ to an earlier version of the manuscript with the title "A model
of wealth accumulation". Sylvain Gibaud (gibaudsylvain@gmail.com) thanks Unit%
\'{e} Mixte de Recherche 5219 for its financial support. J\"{o}rgen Weibull
(jorgen.weibull@hhs.se) thanks the Knut and Alice Wallenberg Research
Foundation, the Agence Nationale de la Recherche, Chaire IDEX
ANR-11-IDEX-0002-02,\ and the Tore Browald and Tom Hedelius Foundation for
financial support.} \\
%EndAName
Department of Economics\\
Stockholm School of Economics\\
\medskip }
\date{\today }
\maketitle

\begin{abstract}
The accumulation of individual fitness or wealth is modelled as a population
game in which pairs of individuals are recurrently and randomly matched to
play a game over a resource. In addition, all individuals have random access
to a constant background resource, and their fitness or wealth depreciates
over time. For brevity we focus on the well-known Hawk-Dove game. In the
base-line model, the probability of winning a fight (that is, when both play
Hawk) is the same for both parties. In an extended version, the individual
with higher current fitness or wealth has a higher probability of winning.
Analytical results are given for the fitness/wealth distribution at any
given time, for the evolution of average fitness/wealth over time, and for
the asymptotics with respect to time and population size. Long-run average
fitness/wealth is non-monotonic in the value of the resource, thus providing
a potential explanation of the curse of the riches.

\bigskip

\end{abstract}

\section{Introduction}

This paper analyzes the\ accumulation and depreciation of personal fitness
or wealth in a finite population where individuals are recurrently and
randomly paired to interact with each other. The interaction takes the form
of a symmetric two-player game over some resource, and the payoffs,
representing gains and losses, are added to and subtracted from the two
individuals' current levels of fitness or wealth. For the sake of
definiteness and brevity, we focus on a simple but canonical Hawk-Dove game.
However, the machinery applies to any symmetric finite game, and can readily
be extended to arbitrary finite games. We generalize a deterministic
mean-field model sketched in Weibull (1999). To the best of our knowledge,
the present study is the first analysis of the stochastic accumulation and
depreciation of individual and average fitness or wealth in populations of
strategically interacting individuals. As will be seen, inequality arises
over time even if all individuals are \textit{ex ante} identical and start
out from the same level of fitness or wealth.

The Hawk-Dove game was used by Maynard Smith and Price (1973) as an
illustration of the possibility that a mixed strategy may be evolutionarily
stable. While they only considered expected payoffs, we here consider
realized payoffs. In particular, if both players use the Hawk strategy, a
fight results in which the winner takes all and the loser incurs a loss.
Consequently, their individual levels of fitness or wealth diverge after
such an interaction. In our base-line model, we follow Maynard Smith and
Price in assuming that all individuals have the same chance of winning a
fight. In an extended version, the more fit or wealthier individual has a
higher chances of winning a fight.

The main focus of this study is on the induced stochastic population
dynamics. While the individuals in our model may be animals or other
biological organisms, and payoffs may be interpreted as increments to
personal fitness, as in Maynard-Smith's and Price's original contribution,
we subsequently interpret the model only in terms of personal wealth, where
the game represent economic opportunities for production or trade,
opportunities that spontaneously arise for given natural resources and
institutions. Whenever such an interaction opportunity arises, each of the
two parties may seek cooperation (play "Dove") or conflict (play "Hawk"). If
both seek cooperation, they split the resource equally. If both seek
conflict, one of them wins the resource and the other individual incurs a
loss. If one individual seeks cooperation and the other conflict, the latter
obtains the full value of the resource and the former neither has a gain or
loss. This is but one example of games that may be used to represent
economic activity in the present framework.

In addition to the game payoffs, individuals now and then receive a constant
background income, and their accumulated fitness wealth is subject to
depreciation over time. This defines an ergodic Markov process. We analyze
this process both in terms of the distribution of individual wealth, and in
terms of average wealth, both at fixed and given times and finite population
sizes, and asymptotically in time and population size. We derive expressions
for long-run average wealth and its variance, and show that this average
agrees with Maynard Smith's and Price's static result (modulo a factor
representing depreciation). However, in the present model, individual wealth
levels are perpetually fluctuating random variables. We illustrate the shape
of the invariant distribution by means of numerical examples.

One of our analytical findings is that average wealth, even in the base-line
model where all individuals have the same chance of winning a fight, is not
monotonically related to the value $v$ of the resource. If the cost of
losing a fight is denoted $c$ (we use the same notation as in Maynard Smith
and Price, 1973), then average fitness or wealth, at any fixed and given
value of $c$, is increasing in $v$ when $v<c/2$, decreasing in $v$ when $%
c/2<v<c$, and (linearly) increasing in $v$ when $v>c$. The reason for this
non-monotonicity is that an increase in the value $v$ inspires individuals
to more fighting, and hence more losses. The model thus provides a possible
explanation for the "curse of the riches".\footnote{%
The `curse of riches' or `curse of natural resources' is an empirical result
from the 1990s that shows a negative correlation, in cross-country studies,
between countries' natural-resource abundance and their economic growth,
after controlling for other relevant variables.} Compared with other
explanations for that phenomenon, the present explanation is perhaps closest
to the rent-seeking explanation, see Torres et al. (2013) for a survey and
discussion.

The model is evidently based on heroic simplifications that permit
analytical results, alongside numerical simulations. We hope that the tools
and methods developed here can be applied to more complex and realistic
models. For surveys of the economics literature on these topics, see
Bardhan, Bowles and Gintis (1999) and Davies and Shorrock (1999), and see
Picketty (2014) and followers for more analysis, discussions and also for
more recent data. In the biology literature, our analysis falls in the
category of "animal fighting",\ a literature pioneered by Enquist and Leimar
(1983, 1984, 1987) and Houston and McNamara (1988).\footnote{%
For a survey of the literature see Hammerstein and Leimar (2015).} The
latter paper analyzes the Hawk-Dove game with an added state variable that
represents the animal's level of energy reserves. Other models of animal
fighting are developed in Crowley (2000) and McNamara and Houston (2005),
briefly discussed in Remark \ref{BioLit} below.

The material is organized as follows. In Section 2 we set up the model, as
applied to a Hawk-Dove game. In Section 3, establish that the wealth process
is ergodic. The larger the population, the less correlated are the wealth
levels within in any finite group of individuals (of fixed size), at any
given time. In the limit as population size tends to infinity, these wealth
levels become statistically independent and identically distributed, a
"propagation of chaos" result. We there solve analytically for the
mean-value and variance of a representative individual's wealth in the
double limit when both time and population size tend to plus infinity.
Section 4 is devoted to the dynamics and asymptotic properties of average
wealth. In particular, the expectation of average wealth follows the
solution to a mean-field differential equation for any population size, and
we show that under drastic depreciation this is also true for average wealth
in a very large population. The mean-field equation is also used to
establish the above-mentioned "curse of the riches" result. Section 5
considers a setting in which the wealthier (or, in the biological context,
more fit) individual in a match has a higher probability of winning a fight.
We there identify the different Nash equilibria, conditional upon current
wealth, establish ergodicity, and show by way of numerical simulations that
both average and median wealth increases. Section 6 concludes. All
mathematical proofs are given in an Appendix at the end of the paper, and
the used simulation programs are available at
https://github.com/ProfesseurGibaud/Model-of-wealth-Accumulation.

\medskip

\section{Model}

\medskip

\label{Model}

Let $\mathbb{N}$ be the set of positive integers, $\mathbb{N}_{0}$ the
nonnegative integers, $2\mathbb{N}$ the even nonnegative numbers, $\mathbb{R}
$ the real numbers, and $\mathbb{R}_{+}$ the nonnegative reals. Consider a
population consisting of a large finite number $N\in \mathbb{N}$ of
individuals who are now and then randomly matched in pairs to play a
Hawk-Dove game (Maynard Smith and Price, 1973). This game, $G\left(
v,c\right) $, is defined by its two positive parameters, the "value" $v$\
and "cost" $c$. Each player has only two pure strategies, H ("hawk") and D
(\textquotedblright dove\textquotedblright ), where the first is
"aggressive" and the second "docile".\ The paired individuals make their
strategy choices simultaneously. If both choose D, they split the value
equally, and thus they each receive payoff $v/2$.\footnote{%
In the biology literature, it is usual to instead assume that one of the
players receives payoff $v$ and the other receives nothing (see e.g.
McNamara and Houston, 2005).} If exactly one of them chooses H, then this
player earns payoff $v$ while the other's wealth is unchanged. If both
choose H, then one wins $v$ and the other loses $c$, the cost of a lost
fight, with equal probability for both. We will call the strategy profile DD 
\textit{cooperation} and the strategy profile HH \textit{fight}. The game is
shown in extensive form in Figure 1 below. The material gains and losses to
player 1 are indicated above those of player 2. If both play H, then
"nature" (player 0) makes a random draw, resulting in a "winner" and a
"loser", with equal chance for both individuals.\footnote{%
By explictly accounting for the outcome of a fight, we depart from the usual
treatment in evolutionary game theory, where expected, not realized, payoffs
are considered.}

\begin{figure}
	\centering
	\includegraphics[width=0.7\linewidth]{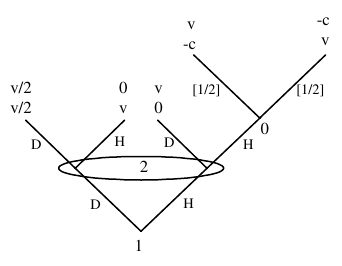}
	\caption{The extensive-form representation of the Hawk-Dove game $G(v,c)$}
	\label{fig:qxb05900}
\end{figure}

Individuals accumulate payoffs over time, depending on how they fare when
playing the Hawk-Dove game with randomly drawn opponents. An individual's
stock of accumulated payoffs at any point in time is called the individual's
current \textit{wealth}. Hence, an individual $i\in \left\{ 1,...,N\right\} $
who enters a pairwise interaction with wealth $w$, exits the interaction
with wealth $w+v/2$ if both play D, with wealth $w+v$ if she plays H and her
opponent plays D, and with unchanged wealth, $w$, if she plays D and the
opponent plays H. If both play H, she will end up with either wealth $w+v$,
which happens with probability one half, or with wealth $w-c$. In terms of
total wealth in the population, all three strategy profiles DD, DH and HD
thus result in an increase by $v$, while the strategy profile HH, to be
called a \emph{fight}, results in a net increase of population wealth by $%
v-c $. For analytical convenience, we\ henceforth assume that $c$\ and $v$\
are positive integers and that $v$\ is even.\footnote{%
Arguably, this assumption is innocuous, since for a sufficiently smallest
unit of wealth, $v$ and $c$ are large and can be arbitrarily well
approximated in this way. The subsequent results hold also when $v$ and $c$
a real numbers, by way of treating the wealth process as a so-called jump
process.}

A pure or mixed strategy in a finite and symmetric two-player game is \emph{%
evolutionarily stable}, or an \emph{ESS} (Maynard Smith and Price, 1973) if
it is a best reply to itself and a better reply to all other best replies.
Hence, it is a refinement of symmetric Nash equilibrium, a refinement that
captures resistance against any "mutant" strategy that appear in a
sufficiently small population share. The Hawk-Dove game has a unique ESS,
namely to use strategy H with probability 
\begin{equation*}
x^{\ast }=\min \left\{ 1,v/c\right\} .
\end{equation*}%
The expected payoff when both parties use this strategy is 
\begin{equation*}
\pi ^{\ast }=\max \left\{ \left( 1-\frac{v}{c}\right) \frac{v}{2},\frac{v-c}{%
2}\right\} .
\end{equation*}

In the present model, the described strategic interaction is not the only
source of wealth. Individuals also have other sources of income. For
analytical convenience, the arrival time of external income is the same as
the arrival time for playing the game. Each party then receives income $%
y/2>c $. This income adds to the gain or loss made in the pairwise game
playing. The assumption $y/2>c$ guarantees that the net gain at each random
match is nonnegative, thus avoiding that an individual's wealth can become
negative. In addition to receiving game payoffs and background income, every
individual's accumulated wealth is exposed to random depreciation, whereby
it stochastically decreases, and may fall down to zero (but not become
negative).\footnote{%
In the biological context, where wealth is replaced by fitness, depreciation
can be interpreted as gradual decay in fitness in the absence of energy
intake, as occasional accidents or illnesses.}

Formally, we model the evolution of wealth for all $N$ individuals as a
Markov process $W^{N}$ in $\mathbb{N}_{0}^{N}$ over continuous time $t\geq 0$%
. Its \textit{state} at any time $t$ is the vector\ $W^{N}(t)=\left(
W_{1}^{N}(t),...,W_{N}^{N}(t)\right) $ of individual wealth holdings. The 
\emph{wealth distribution} at time $t$ is the probability distribution on $%
\mathbb{N}_{0}^{N}$ defined by 
\begin{equation}
\mu ^{N}(t)=\frac{1}{N}\sum_{i=1}^{N}\delta _{W_{i}^{N}(t)}^{\left( i\right)
}\ ,  \label{mu}
\end{equation}%
where $\delta _{W_{i}^{N}(t)}^{\left( i\right) }$ is the probability
distribution on $\mathbb{N}_{0}^{N}$ that assigns unit probability to the
wealth vector where all components $j\neq i$\ are zero and component $i$\
equals $W_{i}^{N}(t)$. The associated \emph{average wealth} is%
\begin{equation}
\bar{W}^{N}(t)=\frac{1}{N}\sum_{i=1}^{N}W_{i}^{N}(t).  \label{av}
\end{equation}

We attach to each individual a \textquotedblleft depreciation Poisson
clock\textquotedblright\ of rate $1$, and to each ordered pair of distinct
individuals a \textquotedblleft game Poisson clock\textquotedblright\ of
rate $1/\left( N-1\right) $. Being statistically independent, the
superposition of these Poisson processes, across the whole population,
results in a stationary population Poisson process, and the accompanying
population wealth process changes state precisely at its arrival times. The
intensity of the aggregate population Poisson process is $\lambda =2N$. To
see this, note that both the \textquotedblleft population depreciation
Poisson clock\textquotedblright\ and the \textquotedblleft population game
Poisson clock\textquotedblright\ each have intensity $N$ (the number of
ordered pairs being $N\left( N-1\right) $).

When the game Poisson clock rings for two individuals, they each receive
income $y/2$ and play the game $G(v,c)$. In this baseline version of the
model, we assume that all individuals always use the unique evolutionarily
stable strategy\textbf{\ }$x^{\ast }=\min \left\{ v/c,1\right\} $ when
playing the game (see Section 5 for a generalization).

When the depreciation Poisson clock rings for an individual, a random
reduction occurs of the individual's wealth. The expected reduction factor
is fixed at $\delta $, and the probability for reduction to zero wealth has
a uniform positive lower bound $\varepsilon $. More precisely, let $\delta
\in \left[ 0,1\right] $ and $\varepsilon >0$. For each $k\in \mathbb{N}$,
let $Z_{k}$ be a random variable that takes values in $\left\{
0,1,...,k\right\} $, with $\mathbb{E}\left( Z_{k}\right) =\delta k$ and $%
\mathbb{P}\left( Z_{k}=k\right) \geq \varepsilon $. If $W_{i}^{N}(t)=w>0$
then depreciation replaces the individual's current wealth $w$ by the random
wealth level $w-Z_{w}$ (with statistical independence with respect to all
other random events). If $W_{i}^{N}(t)=0$, then the individual's wealth
after depreciation remains at zero. Hence, the conditionally expected
individual wealth level, after depreciation, given current wealth $%
W_{i}^{N}(t)=w>0$, is $\left( 1-\delta \right) w$, and the probability for
losing all wealth has a positive lower bound, $\varepsilon $, for all $w>0$.
We call\textbf{\ }$\delta $\textbf{\ }the \emph{depreciation rate}.

The described events of pairwise matching for game play and individual
wealth depreciation are all statistically independent. Given the Poisson
clocks, the underlying game $G\left( v,c\right) $, the associated ESS $%
x^{\ast }$, the depreciation rate $\delta $, and probability bound $%
\varepsilon $, $W^{N}=\left\langle W^{N}(t)\right\rangle _{t\in \mathbb{R}%
_{+}}$ constitutes a Markov process in $\mathbb{N}_{0}^{N}$.

\medskip

\section{Asymptotics}

\medskip

\label{Prop du chaos, Jorgen Weibull}

We here provide two asymptotic results for the wealth process. First with
respect to time, for a given finite population, and then with respect to
population size, at a given finite observation time.

Consider a population of arbitrary fixed size, $N>1$.\ At any finite time $t$%
, the state of the wealth process is evidently history dependent. However,
asymptotically over time it is not. The effect of the initial wealth
distribution washes out over time an vanishes asymptotically.\textbf{\ }This
is not surprising, given the nature of the process. However, to formally
prove this is non-trivial, so a proof is given in the appendix for the
interested reader.

\medskip

\begin{proposition}
\label{TE}The wealth process $W^{N}$ is ergodic, and thus has a unique
invariant distribution to which it converges in distribution from any
initial state.
\end{proposition}

\medskip

Second, we instead consider the distribution of individual wealth at any
fixed and given time $t$\ when the population is very large. To be more
specific, we analyze the probability distribution for any given individual's
wealth\ at a given time $t>0$\textbf{\ }in the limit as $N\rightarrow
+\infty $. For this purpose, let $\mathcal{L}(X)$ denote the probability
distribution of any random variable $X$, and let $\left( \mathcal{L}%
(X_{i})\right) ^{\otimes k}$ be the product probability distribution of $k$
such i.i.d. random variables $X_{i}$.

Suppose that the initial individual wealth levels, the random variables $%
W_{i}^{N}(0)$, for $i=1,...,N$, are i.i.d. $\pi $, irrespective of
population size $N$. Then one can establish the following "propagation of
chaos" result for the arguably most interesting case when the Hawk-Dove game
is not dominance solvable, that is, when $v<c$.

\medskip

\begin{proposition}
\label{T2} Suppose that $v<c$ and that all individuals always use strategy $%
x^{\ast }$. For any initial probability distribution $\pi $, the evolution
of individual wealth can be represented by a Markov process\ $\tilde{W}%
=\left\langle \tilde{W}(t)\right\rangle _{t\in \mathbb{R}_{+}}$ in $\mathbb{N%
}_{0}$, with $\mathcal{L}(\tilde{W}(0))=\pi $, such that for any number $%
n\in \mathbb{N}$ of individuals: 
\begin{equation}
\lim\limits_{N\rightarrow +\infty }\mathcal{L}(W_{1}^{N},\dots
,W_{n}^{N})=\left( \mathcal{L}(\tilde{W})\right) ^{\otimes n}.  \label{conv}
\end{equation}%
Moreover, when $\mathbb{P}\left( Z_{k}\in \left\{ 0,k\right\} \right) =1$,
then the following "propagation of chaos" differential equation holds, for
any\textbf{\ }$w\in \mathbb{N}_{0}$ and $t\geq 0$: 
\begin{eqnarray}
\frac{\partial }{\partial t}\mathbb{P}\left[ \tilde{W}\left( t\right) =w%
\right] \  &=&\ 2\left( 1-\frac{v}{c}\right) ^{2}\cdot \mathbb{P}\left[ 
\tilde{W}\left( t\right) =w-v/2-y/2\right]   \label{eqn Evol Eq Jorgen} \\
&&+\frac{2v}{c}\left( 1-\frac{v}{c}\right) \cdot \left( \mathbb{P}\left[ 
\tilde{W}\left( t\right) =w-v-y/2\right] +\mathbb{P}\left[ \tilde{W}\left(
t\right) =w-y/2\right] \right)   \notag \\
&&+\left( \frac{v}{c}\right) ^{2}\cdot \left( \mathbb{P}\left[ \tilde{W}%
\left( t\right) =w-v-y/2\right] +\mathbb{P}\left[ \tilde{W}\left( t\right)
=w+c-y/2\right] \right)   \notag \\
&&-2\cdot \mathbb{P}\left[ \tilde{W}\left( t\right) =w\right] -D_{w}  \notag
\end{eqnarray}%
where $D_{w}=\delta \cdot \mathbb{P}\left[ \tilde{W}\left( t\right) =w\right]
\ $for all $w\neq 0$, and\textbf{\ }$D_{0}=\delta \cdot \mathbb{P}\left[ 
\tilde{W}\left( t\right) =0\right] -\delta $.
\end{proposition}

\medskip

The process $\tilde{W}$ can be thought of as the wealth dynamics of a 
\textit{representative individual}. The first part of this theorem, the
convergence result\ (\ref{conv}), establishes that the larger the
population, the less correlated are the wealth levels within in any finite
group of individuals (of fixed size $k$), and, in the limit as population
size $N$ tends to infinity, these individual wealth levels become
statistically independent. Moreover, the probability distribution of each
individual's wealth, $W_{i}^{N}\left( t\right) $, at any given time $t>0$,
tends to the distribution of the random variable $\tilde{W}\ $as population
size $N$ tends to infinity.

The evolution of this probability distribution over time $t$ is given in
equation (\ref{eqn Evol Eq Jorgen}), for the special case when depreciation
takes the drastic form of either leaving the individual's wealth untouched
(with probability $1-\delta $) or making it vanish altogether (with
probability $\delta $). Technically, the assumption is that for an
individual with wealth $w$, the random wealth loss, $Z_{w}$\ takes values in 
$\left\{ 0,w\right\} $, that is, either all $w$\ units vanish, or the
individual's wealth remains intact.\footnote{%
This assumption is made only in order for the chaos evolution equations to
be simple.} Moreover, for any level of wealth $w$, $\mathbb{P}(\tilde{W}%
(t)=w)$ is the population share of individuals, in an infinite population,
with that wealth level. The different terms in the evolution equation
represent different "inflows" and "outflows" from any given wealth level $%
w\in \mathbb{N}_{0}$.\ More precisely, there are three inflows, from wealth
levels $w-v/2$,$\ w-v\ $and$\ w+c$, and two outflows, one because
individuals are drawn to play the game and one because of depreciation of
wealth. {The coefficients in equation (\ref{eqn Evol Eq Jorgen}) can be
obtained from Figure 1 by multiplying probabilities downwards from the
terminal nodes to the root of the tree, under the hypothesis that individual 
}$i${\ at the root has wealth }$w${, and using the fact that the average
time rate of game-playing for an individual }is always 2, also{\ in the
limit as }$N\rightarrow \infty ${. The} wealth level zero, however, is
special, if $\delta >0$, in that it has no depreciation outflow. Instead, it
has an extra inflow, emanating from depreciation of the wealth of
individuals with non-zero wealth.\footnote{%
A more realistic modelling of depreciation would not have this feature.
However, such a richer model would not allow the analytical tractability we
now have.}

One may use equation (\ref{eqn Evol Eq Jorgen}) to derive the laws of motion
for the first and second moments of the distribution of a representative
individual's wealth at any point in time, granted the depreciation rate is
positive.

\begin{corollary}
\label{C2} Suppose that (a) $v<c$, (b) depreciation is such that $\mathbb{P}%
\left( Z_{k}\in \left\{ 0,k\right\} \right) =1$, (c) all individuals always
use strategy $x^{\ast }$, and (d) the initial individual wealth distribution 
$\pi $ has finite mean and variance. There then exist constants $%
K_{1},K_{2},K_{3},K_{4}\in \mathbb{R}$ such that%
\begin{equation}
\mathbb{E}\left[ \tilde{W}(t)\right] \ =\ \frac{1}{\delta }\left[ \left( 1-%
\frac{v}{c}\right) v+y\right] +K_{1}e^{-\delta t}\quad \forall t\geq 0
\label{m1}
\end{equation}%
\begin{equation}
Var\left[ \tilde{W}(t)\right] \ =\ \frac{1}{\delta ^{2}}\left[ \delta
A+\left( 1-\frac{v}{c}\right) ^{2}v^{2}+y^{2}-2\frac{v}{c}vy\right]
+K_{2}e^{-\delta t}+K_{3}e^{-2\delta t}+K_{4}te^{-\delta t}\quad \forall
t\geq 0.  \label{m2}
\end{equation}%
for%
\begin{equation*}
A=\frac{1}{2}\left[ 4-\left( 1-\frac{v}{c}\right) ^{2}\right] v^{2}+\frac{1}{%
2}\left[ \left( 2-\frac{v}{c}\right) \frac{v}{y}+1\right] y^{2}.
\end{equation*}
\end{corollary}

\bigskip

In other words, the first and second moments converge exponentially over
time to their asymptotic values, 
\begin{equation}
\lim_{t\rightarrow \infty }\mathbb{E}\left[ \tilde{W}(t)\right] =\frac{1}{%
\delta }\left[ \left( 1-\frac{v}{c}\right) v+y\right]  \label{EW}
\end{equation}%
and%
\begin{equation}
\lim_{t\rightarrow \infty }Var\left[ \tilde{W}(t)\right] =\frac{1}{\delta
^{2}}\left( \delta A+\left( 1-\frac{v}{c}\right) ^{2}v^{2}+y^{2}-2\frac{y}{c}%
v^{2}\right)  \label{VarW}
\end{equation}

It follows that the asymptotic \textit{coefficient of variation} (also known
as "relative standard deviation"), is decreasing in the depreciation rate $%
\delta $\ and, as the this rate $\delta $\ tends zero, the coefficient of
variation tends to 
\begin{equation}
\lim_{\delta \rightarrow 0}\lim_{t\rightarrow +\infty }\frac{\sqrt{Var(%
\tilde{W}(t))}}{\mathbb{E}(\tilde{W}(t))}=\frac{\sqrt{\left[ \left( 1-\frac{v%
}{c}\right) v+y\right] ^{2}-2yv}}{\left( 1-\frac{v}{c}\right) v+y}
\label{CV}
\end{equation}

The\ variation of an individual's wealth has three sources. First the
stochastically arriving income $y/2$. Second, the strategic interaction,
which gives rise to gains and losses, a source represented by the game
parameters $v$\ and $c$, where $v<c$. Third, the depreciation of individual
wealth, a source\ represented by the parameter $\delta $, here taken down to
zero. We note that the limit coefficient of variation, when $\delta
\downarrow 0$, is always less than one, and that it equals one in the
absence of the exogenous income $y$, for any values $v<c$. We conjecture
that this invariance is due to the equilibrium nature of individual
behavior, the probability $x^{\ast }$. For both pure strategies, H and D,
then result in the same expected net wealth gain.

The wealth distribution is different when the depreciation hits less
dramatically than assumed in the above calculations. For example, suppose
that depreciation can take two forms, where the first is the same as above,
that is, that an individual's all wealth vanishes with probability $\delta $%
, while the second form is that each unit of an individual's wealth vanishes
with the same probability $\delta $\ and that this event is statistically
independent among wealth units. Suppose moreover, that the first form of
depreciation has probability $\rho $\ and the second has probability $1-\rho 
$, and assume that all these random variables are statistically independent.
Hence, in the second case, the number $Z_{w}$\ of depreciated wealth units
is binomially distributed, $Bin\left( w,\delta \right) $. See diagram below.

\begin{figure}[h!]
	\centering
	\includegraphics[width=1\linewidth]{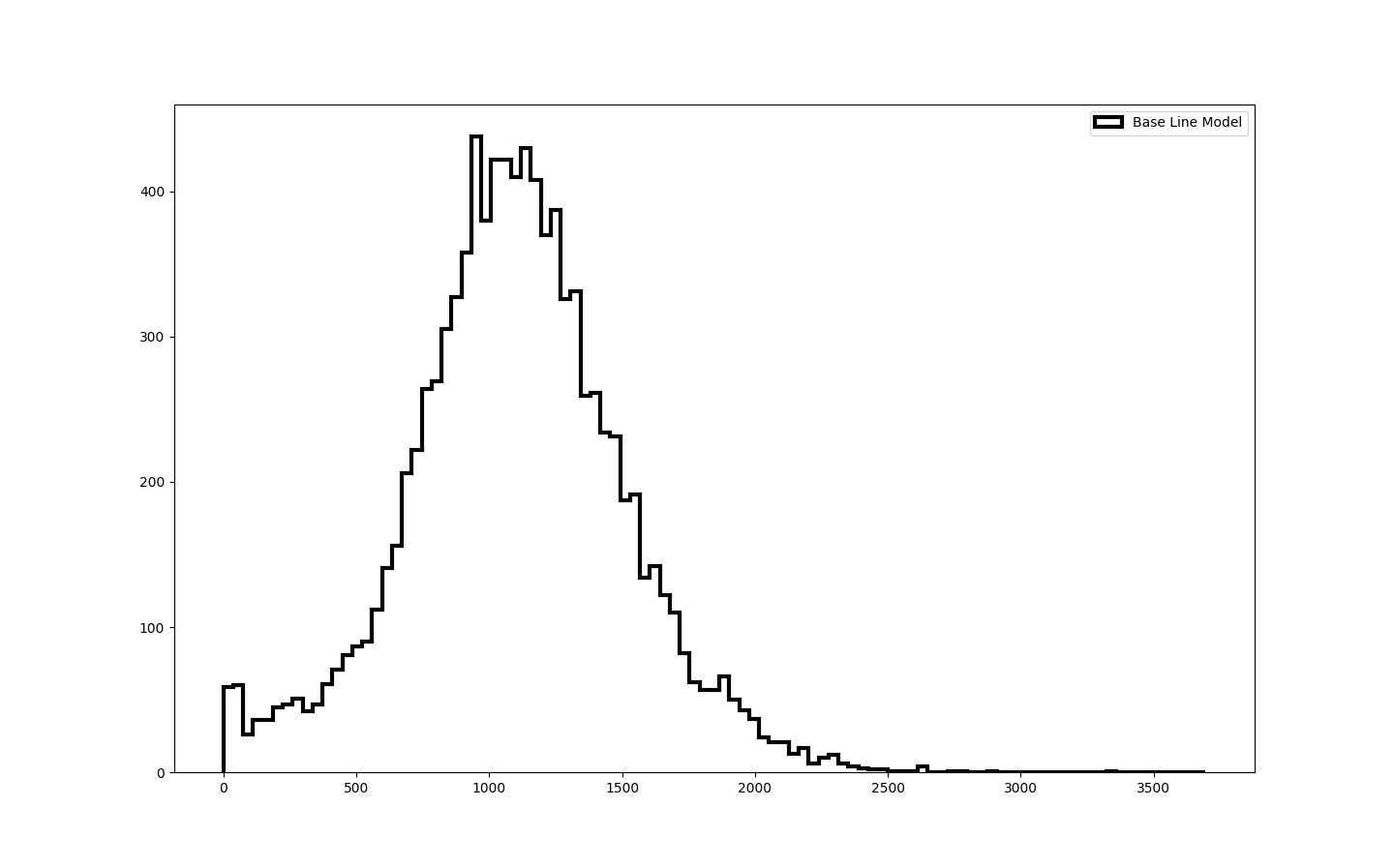}
	\caption{Empirical long-run wealth distribution under partly binomial depreciation.}
	\label{fig:210710versiondfigure3}
\end{figure}

The figure shows simulation results for $v=40$, $c=20$, $y=100$, $\delta
=0.1 $, $N=10,000$, $T=10,000,000$, and $\rho =1/10$. In this simulation,
the empirical average (with respect to a given individual) is approximately
1098, the standard deviation is 403, the median 1095, the maximum wealth
3347, and the Gini coefficient 0.2. We note that the empirical coefficient
of variation thus is approximately 0.37, while the theoretical asymptotic
value for $\rho =1$, according to (\ref{CV}), is approximately 0.44. As
expected, the less drastic version of depreciation induces a lower
coefficient of variation.

\section{Average wealth}

\medskip

Let $T$ be any arrival time of the above Poisson process and let $\bar{W}%
^{N}(T)$ be average wealth in the population at this time, defined in (\ref%
{av}). Hence, "population wealth, or "national wealth" is $N\bar{W}^{N}(T)$.
At this arrival time $T$, one of two equally probable events will take
place: either one individual is selected for wealth depreciation, or an
ordered pair of individuals are selected to receive income and play the game 
$G\left( v,c\right) $.

In the first event, the wealth of one randomly drawn individual is taken to
zero with probability $\delta $. Accordingly, average wealth in the
population then decreases by a random integer amount $Y^{N}$ (depending on
the wealth of the randomly drawn individual), the expected value of which,
conditional upon $\bar{W}^{N}(T)$, is $\delta \cdot \bar{W}^{N}(T)/N$. In
the second event, there is a positive probability that the two matched
individuals will fight, that is, both play H, and this probability is $%
\left( x^{\ast }\right) ^{2}=\left( \min \left\{ 1,v/c\right\} \right) ^{2}$%
. When this happens, average wealth will change by $\left( v-c\right) /N$.
If they will not both fight, then no wealth is lost in their interaction,
and average wealth increases by $v/N$.

In sum, average wealth at all times $t\in \lbrack T,T^{\prime })$, where $%
T^{\prime }$ is the next arrival time, is determined as follows:%
\begin{equation}
\bar{W}^{N}(t)\quad =\quad \bar{W}^{N}(T)\ +\ \frac{1}{N}\cdot \left\{ 
\begin{array}{ll}
-Y^{N} & \text{{\small with probability}\textsf{\ }}\frac{1}{2} \\ 
v-c+y & \text{{\small with probability}\textsf{\ }}\frac{1}{2}\left( \min
\left\{ 1,v/c\right\} \right) ^{2} \\ 
v+y & \text{{\small otherwise}}%
\end{array}%
\right. .  \label{w}
\end{equation}

This defines a stochastic process $\bar{W}^{N}$ that, however, is not a
Markov process. The reason is that when the population depreciation Poisson
clock strikes, the statistical distribution of depreciation depends on the
current wealth distribution. Average wealth falls more if a rich individual,
instead of a poor individual, is hit by "drastic depreciation".

\subsection{The case $v<c$\label{case v<c}}

In this classical case in evolutionary game theory, the unique
evolutionarily stable strategy in $G\left( v,c\right) $ is to play H with
probability $x^{\ast }=v/c$. If everybody else in the population uses this
strategy, then it is an optimal strategy for any individual who strives to
maximize\ his or her expected net wealth gain in each interaction. If all
individuals play $x^{\ast }=v/c$ in all matchings, then the probability for
a fight (HH) in a random match is $q^{\ast }=v^{2}/c^{2}$.

One would guess that when $N$\ is large, then the average-wealth process
follows closely the solution trajectories of its mean-field equation. To
make this precise, suppose that all individuals in all matches play the
unique ESS, $x^{\ast }=v/c$. Taking expectations in (\ref{w}) suggests a
simple time-homogeneous ordinary differential equation for the dynamics of
expected average wealth, from any initial state.

\begin{proposition}
\label{mean field}Let $w(t)=\mathbb{E}\left[ \bar{W}^{N}\left( t\right) \mid 
\bar{W}^{N}\left( 0\right) =w_{0}\right] $, and suppose that $\delta >0$.
Then 
\begin{equation}
\dot{w}(t)=v\left( 1-\frac{v}{c}\right) +y-\delta w(t)\quad \forall t\geq 0,
\label{ODE}
\end{equation}%
with initial value $w\left( 0\right) =w_{0}$.
\end{proposition}

In the Appendix we prove this by using infinitesimal generators. This simple
mean-field equation has a unique solution, namely, 
\begin{equation}
w(t)=\left( 1-\frac{v}{c}\right) \frac{v}{\delta }+\frac{y}{\delta }+\left(
w_{0}-\left( 1-\frac{v}{c}\right) \frac{v}{\delta }-\frac{y}{\delta }\right)
e^{-\delta t}\mathsf{\quad }\forall t\geq 0\text{.}  \label{sol}
\end{equation}%
Irrespective of the initial wealth level $w_{0}\geq 0$, this solution
converges asymptotically to the unique steady-state level (see Corollary \ref%
{C2}) 
\begin{equation}
w^{\ast }=\left( 1-\frac{v}{c}\right) \frac{v}{\delta }+\frac{y}{\delta }%
=\lim_{t\rightarrow \infty }\mathbb{E}\left[ \tilde{W}(t)\right] .
\label{wlim1}
\end{equation}

We also note that average wealth increases\ linearly over time in the
absence of depreciation: for $\delta =0$, the solution to (\ref{ODE}) is{%
\begin{equation}
w(t)=w_{0}+yt+v\left( 1-\frac{v}{c}\right) t\ \quad \forall t\geq 0,
\label{wt}
\end{equation}%
}for any initial wealth level {$w_{0}{\geq 0}$.\newline
}\medskip

\subsection{The case $v>c$}

\bigskip

What happens if the opportunity value $v$ exceeds the damage cost $c$? As
noted above, it is then always optimal to play strategy H. Suppose that this
is what all individuals do. Then the the mean-field equation for average
wealth becomes 
\begin{equation}
\dot{w}(t)=y+v-c-\delta w(t).  \label{wsol2}
\end{equation}%
Accordingly, all solutions, irrespective of initial conditions, converge to
the steady state level 
\begin{equation}
w^{\ast }=\frac{y+v-c}{\delta }.  \label{wlim2}
\end{equation}

The above approximation result, Proposition \ref{mean field}, holds as
stated with equation (\ref{ODE}) replaced by equation (\ref{wsol2}). {Also
in this case average wealth} would increase linearly over time in the
absence of depreciation:%
\begin{equation}
w(t)=w_{0}+\left( y+v-c\right) t\ \quad \forall t\geq 0.  \label{wlin}
\end{equation}

\medskip

\subsection{Comparative statics}

\medskip

Combining equations (\ref{wlim1}) and (\ref{wlim2}) we obtain the following
general expression for the unique steady-state level of average wealth,
associated with any positive depreciation rate $\delta $:%
\begin{equation}
w^{\ast }=\max \left\{ \frac{y+v-v^{2}/c}{\delta },\frac{y+v-c}{\delta }%
\right\} .  \label{w*}
\end{equation}

Not surprisingly, average wealth is lower the higher is the depreciation
rate. However, the equation also shows a feature that may be less expected,
namely, that steady-state average wealth is non-monotonic in both the value $%
v$ of the stake in the pairwise opportunities and in the cost $c$ of a lost
conflict. This is illustrated in the two diagrams below.\ Figure 3 shows
steady-state average wealth as a function of $v$, for $c=40$, $y=100$ and $%
\delta =0.1$. The dashed horizontal line indicates steady-state wealth in
the absence of the pairwise interaction. Figure 4 again shows steady-state
wealth, but now as a function of $c$, for $v=20$, $y=100$ and $\delta =0.1$. 
\pagebreak

\begin{figure}[h!]
	\centering
	\includegraphics[width=0.7\linewidth]{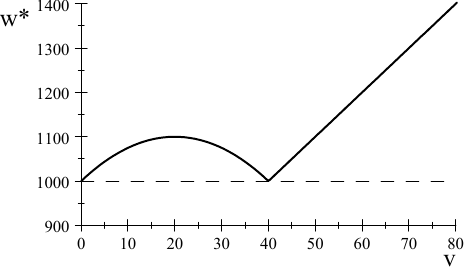}
	\caption{Steady-state average wealth as a function of the value $v$ in game $G(v,c)$.}
	\label{fig:qxb05901}
\end{figure}

\begin{figure}[h!]
	\centering
	\includegraphics[width=0.7\linewidth]{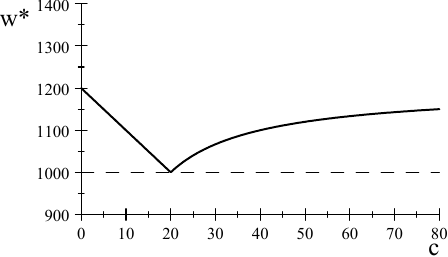}
	\caption{Steady-state average wealth as a function of the cost $c$ in game $G(v,c)$.}
	\label{fig:qxb05902}
\end{figure}

The reason why the game adds nothing to the steady-state wealth level $%
w^{\ast }$ when $v=c$ is that then all matched pairs have a fight, whereby
one party wins $v$ and the other loses just as much. Consequently, for such
game parameters, average wealth converges over time to the steady-state
level $y/\delta =1000$, from any initial level. The reason why steady-state
wealth is also non-monotonic in the cost $c$ of losing a conflict, is the
other side of the same coin. For $c<v$, H dominates D, and thus all pairs
fight, and each time the net gain is $v-c$, a gain that decreases linearly
in $c$. When $c$ has risen to $v$, the net gain is zero, and for higher $c$,
the probability for a fight\ is $v^{2}/c^{2}$, a decreasing function of $c$,
so the expected net gain in a random match is $v-v^{2}/c$, an increasing
function of $c$. The more damage an individual who loses a conflict suffers,
the fewer conflicts there are in equilibrium, and the wealthier will society
be in steady state. Hence, contrary to what one might first think, a
reduction of the damage due to a fight\ may increase the frequency of
fights\ enough to reduce steady-state wealth.

\medskip

\section{When wealth is strength}

\bigskip

So far, we have assumed that all individuals have the same chance of winning
a fight. Arguably, wealthier individuals usually have a higher probability
of winning conflicts. In the animal kingdom, "wealth" may be body weight,
muscular mass, or control of a good territory, while among humans, wealth
may consist in part in buildings and weaponry, or availability of good
lawyers. We here briefly outline how our base-line model can be generalized
to allow for "wealth is strength".

Consider the slight generalization of the (symmetric) HD game $G\left(
v,c\right) $ to the (potentially asymmetric) HD game $G^{\ast }\left(
v,c,p\right) $ shown in the diagram below, where $p\in \left[ 0,1\right] $
is the (exogenous) probability that player 1 will win a fight. We note that
the original Hawk-Dove game is the special case when $p=1/2$.
\begin{figure}
	\centering
	\includegraphics[width=0.7\linewidth]{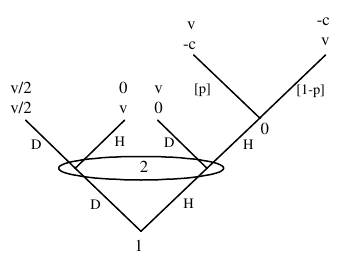}
	\caption{The slightly generalized Hawk-Dove game $G^*(v,c,p)$.}
	\label{fig:qxb05903}
\end{figure}

For brevity, we henceforth focus on the case when $v<c$.

Consider two individuals in the population, $i$\ and $j$, who just have been
matched to play the extended HD game $G\left( v,c,p\right) $, where player
1's success probability $p$ depends on their wealth levels. More precisely,
if individual $i$ has wealth $w_{i}$ and is in player role 1, and individual 
$j$ has wealth $w_{j}$\ and is in player role 2, then $p=f(w_{i},w_{j})$,
where $f:\mathbb{R}^{2}\rightarrow \left[ 0,1\right] $ is the logistic
version of Tullock's contest function (Tullock (1980)): 
\begin{equation}
f(w_{i},w_{j})=\frac{e^{\sigma w_{i}}}{e^{\sigma w_{i}}+e^{\sigma w_{j}}}.
\label{Tull}
\end{equation}%
Here $\sigma \geq 0$\ is a parameter that represents the sensitivity of the
success probability $p$ to the wealth difference between the fighters. It is
increasing in own wealth, $w_{i}$, decreasing in the opponent's wealth, $%
w_{j}$ and equals one half for any value of $\sigma $, when $w_{i}=w_{j}$.
The success probability is also one half when $\sigma =0$, irrespective of
the wealth levels (just as in game $G\left( v,c\right) $). We also note that
it is immaterial which player role, 1 or 2, that individuals are assigned to
(if $i$ is assigned player role 2, then $p=f(w_{j},w_{i})=1-f\left(
w_{i},w_{j}\right) $).

A\ number of relevant information scenarios open up. In one scenario, each
individual only knows his or her own wealth. In another scenario, any two
matched individuals observe each others' wealth. In a third scenario, each
individual in a match knows her own wealth and receives a noisy private
signal about the opponent's wealth. We here focus on the second scenario,
for $\sigma >0$.\footnote{%
For analyses of asymmetric information, see see Enquist and Leimar (1983).
For analyses of asymmetric contest games, see Franke, Kanzow and Leininger
(2013) and references therein.}

It is easily verified that strategy $H$\ strictly dominates strategy $D$ for
individual $i$ (irrespective of player role) if and only if his or her
winning probability is sufficiently high, $f(w_{i},w_{j})>c/\left(
v+c\right) $, a condition that is equivalent with%
\begin{equation}
w_{i}-w_{j}>\frac{1}{\sigma }\ln \left( \frac{c}{v}\right) .  \label{diff1}
\end{equation}

The set of Nash equilibria of $G\left( v,c,p\right) $, when played between
individual $i$\ in player role 1 and individual $j$\ in player role 2,
depends on the parameters as follows (see Appendix for a proof):

\begin{proposition}
\label{Prop strength}Suppose that $v<c$. If (\ref{diff1}) holds, then the
unique Nash equilibrium is $(H,D)$. If 
\begin{equation}
\left\vert w_{i}-w_{j}\right\vert <\frac{1}{\sigma }\ln \left( \frac{c}{v}%
\right) ,  \label{int}
\end{equation}%
then there are three Nash equilibria: $(H,D)$, $(D,H)$, and a mixed
equilibrium in which individual $i$ plays $H$ with probability 
\begin{equation}
x^{\ast }=\frac{v}{2\left( v+c\right) f(w_{i},w_{j})-v}  \label{asym}
\end{equation}%
and individual $j$\ plays $H$\ with probability 
\begin{equation}
y^{\ast }=\frac{v}{2\left( v+c\right) f(w_{j},w_{i})-v}.  \label{asym2}
\end{equation}
\end{proposition}

In sum: when wealth levels are sufficiently apart, the poorer individual
plays D and the richer plays H. The rich individual then takes the whole
"cake" without fight. When wealth levels are not far apart, but not
identical, there are three candidate equilibria. In one equilibrium, the
rich individual takes the cake without fight, in another, the poor
individual takes the cake without fight, and in the third equilibrium they
both randomize between H and D, but with slightly different probabilities.
Irrespective of which equilibrium is played in such encounters, the general
level of fighting in the population is lower than in the base-line Hawk-Dove
model. This conclusion is in line with McNamara and Houston, 2005. (However,
a difference between our model and theirs is that in their model, the two
contestants don't know the other contestant's fighting ability). If the
wealth levels happen to be identical, then we are back to the base-line
model.\footnote{%
We neglect the knife-edge case when $\left\vert w_{i}-w_{j}\right\vert
=\sigma ^{-1}\ln \left( c/v\right) $, since this happens only in very
special circumstances.}

\begin{remark}
\label{BioLit}It is well-known in biology that animals who contest a
resource many times avoid fighting, and thereby avoid damage, by way of
judging each other's strength. Contestants also often try to impress each
other by demonstrating or exaggerating their fighting ability. Usually,
fights occur only if the two contestants appear approximately equally
strong. Arguably, avoidance of conflicts between unequal individuals are
also common among humans. For mathematical models of animal fighting, see
Enquist and Leimar (1983, 1984, 1987, 1990), Houston and McNamara (1988),
Crowley (2000), and McNamara and Houston (2005).\ In Houston and McNamara
(1988), individuals have different `energy reserve' levels, and the ESS has
the form `play H if your energy reserves are below a certain critical value,
otherwise play D'. Crowley (2000) analyze both the case when individuals
only know their own fighting ability and when they also know their
opponent's ability. In McNamara and Houston (2005), where each individual
knows only his or her own fighting ability, the ESS takes the form of a
threshold for switching from D to H as own fighting ability rises. None of
the mentioned studies analyzes the associated stochastic accumulation
processes.
\end{remark}

A possible scenario is that the mixed equilibrium is played whenever (\ref%
{int}) holds; and this is in line with the empirical observation that
fighting seems to take place mostly between parties that are relatively
equal in strength. Formally, the equilibrium strategy for an individual in
the population game (where all individuals are players who are randomly
called upon for pairwise play) can then be defined as a function $\xi $ of
own wealth, $w$, and the opponent's wealth, $w^{\prime }$:%
\begin{equation*}
\xi \left( w,w^{\prime }\right) =\mathbb{P}\left( H\mid w,w^{\prime }\right)
=\left\{ 
\begin{array}{ll}
\quad 1 & \text{if }w-w^{\prime }>\sigma ^{-1}\ln \left( c/v\right) \\ 
\frac{v}{2\left( v+c\right) f(w,w^{\prime })-v} & \text{if }\left\vert
w-w^{\prime }\right\vert \leq \sigma ^{-1}\ln \left( c/v\right) \\ 
\quad 0 & \text{if }w^{\prime }-w>\sigma ^{-1}\ln \left( c/v\right)%
\end{array}%
\right.
\end{equation*}

However, as shown in Selten (1980), mixed equilibria are not played in
evolutionarily stable role-conditioned strategies. Hence, evolutionary
stability requires that either $(H,D)$ or $(D,H)$ is played when $%
0<\left\vert w-w^{\prime }\right\vert \leq \sigma ^{-1}\ln \left( c/v\right) 
$. The following population-game strategy is evolutionarily stable:%
\begin{equation*}
\xi ^{\ast }\left( w,w^{\prime }\right) =\mathbb{P}\left( H\mid w,w^{\prime
}\right) =\left\{ 
\begin{array}{ll}
\quad 1 & \text{if }w>w^{\prime } \\ 
v/c & \text{if }w=w^{\prime } \\ 
\quad 0 & \text{if }w^{\prime }<w%
\end{array}%
\right.
\end{equation*}

Figure 6 below compares the long-run wealth distribution for $\sigma =1$,
under population strategy $\xi ^{\ast }$, with the base line-case when $%
\sigma =0$, for the same parameter values as in Figure 2 (but with slightly
different resolution). Not surprisingly, the distribution is more dispersed
and average wealth is somewhat higher (because of the avoidance of fights).
The empirical average in this simulation is $1206$, standard deviation $506$%
, median $1177$, maximum $3625$, and the Gini coefficient is $0.23$.
Compared with the base line model, with equal probability of winning a
fight, the empirical average, median, maximum increased almost $10$\%, the
Gini coefficient by $15$\%, and the standard deviation by about $25$\%.%
\footnote{%
Under population strategy $\xi $, the empirical average wealth level is
1195, standard deviation 508, median 1171, maximum 3696, and the Gini
coefficient is 0.24. Hence, it does not matter much which population
strategy is used when wealth levels are close.}

\begin{figure}
	\centering
	\includegraphics[width=1\linewidth]{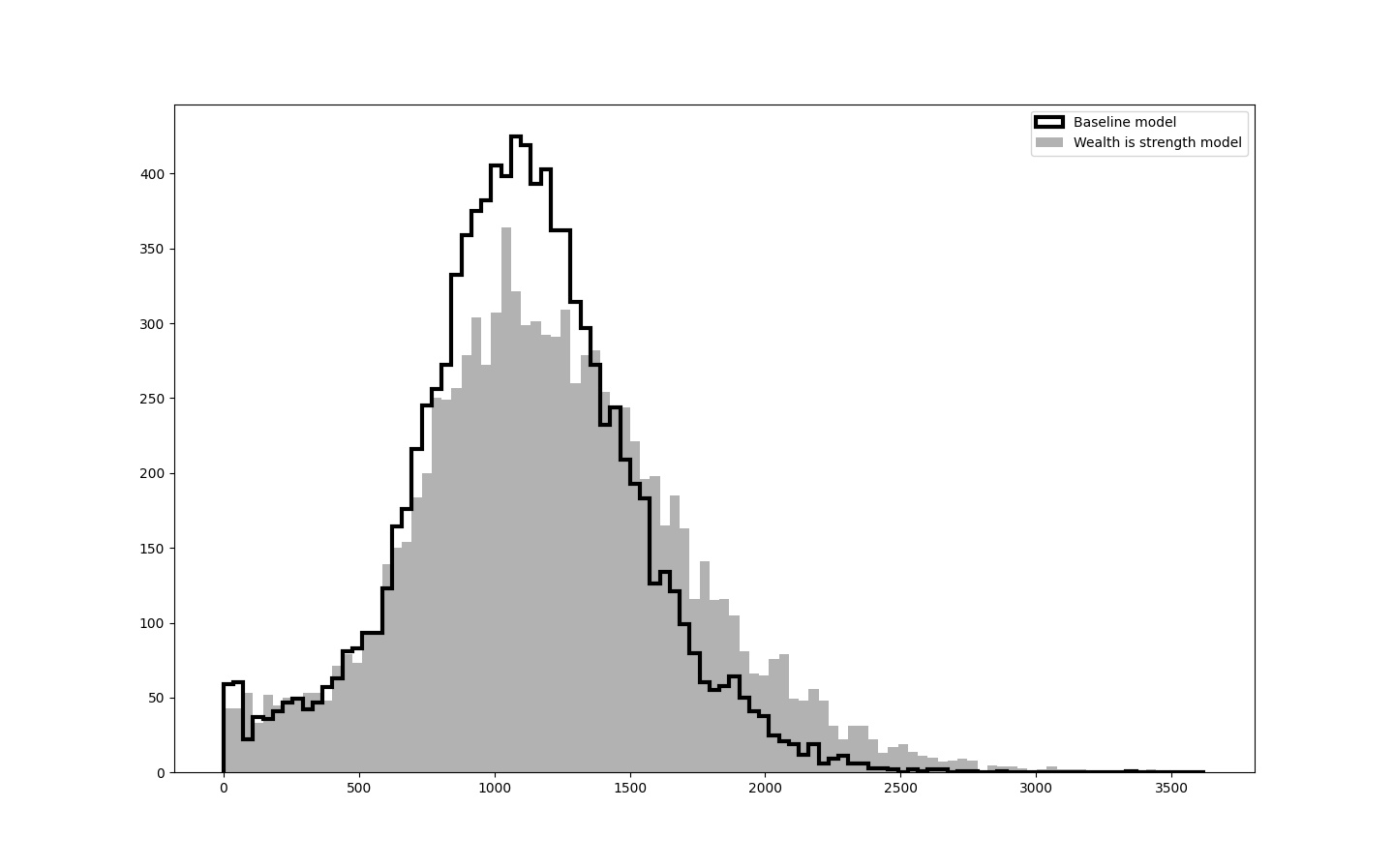}
	\caption{The long-run distribution of individual wealth when "wealth is strength", compared with the distribution in the base-line model.}
	\label{fig:wealth210803figure6versionb}
\end{figure}

When individuals' strategies are state-dependent, as in the present model of
"wealth is strength", some previous results for the base line model still
hold. In particular, the proof of Proposition 1 doesn't change when wealth
is strength. Hence, the wealth process is still ergodic. The fact that
individuals know their own wealth and the wealth of their opponent brings
non-linearity into the proof of the propagation of chaos. For a proof of
propagation-of-chaos for a similar but non-linear wealth accumulation
process, see Gibaud (2016). Based upon that proof, we claim that the
propagation of chaos holds also when wealth is strength. However, since the
evolution equations change, so will Corollary 1.

\textbf{\medskip }

\section{Discussion}

\medskip

The purpose of this study was to work out an analytical framework that
permits rigorous mathematical analysis of mechanisms at work in the
accumulation and distribution of wealth, a framework that would allow
extensions and generalizations to richer and more realistic models.

For instance, instead of having only one game played---here a simple
Hawk-Dove game---there could be a family of more or less complex $n$-player
games (for $n=1,2,3...$) representing opportunities for production, trade,
bargaining etc., games that are randomly drawn according to some (exogenous
or endogenous) probability distribution.\footnote{%
A Hawk-Dove game can be thought of as a simple bargaining game in which H
represents an aggressive claim (the whole cake) and D a modest claim
(splitting the cake equally). Molander (2014) discusses how even slight
differences in bargaining power may induce wide wealth dispersion.}
Individuals may then be given the option of not taking part in an
interaction, which can easily be obtained by adding a pure strategy, that if
chosen by a participant, leaves that individual's wealth untouched (and that
affects other participants' material payoffs in a prescribed way).\footnote{%
In the\ Hawk-Dove game, this option is of little interest, since abstention
would be weakly dominated by strategy D.} We hope that the present
analytical framework, in suitably extended forms, may help understand
mechanisms behind the accumulation and distribution of wealth, see e.g.
Bardhan, Bowles and Gintis (1999), Davies and Shorrock (1999), and Picketty
(2014).

The present analysis rests upon other heroic and unrealistic assumptions. An
important ingredient that is missing in the present model framework is
consumption. What we here call \textquotedblleft
depreciation\textquotedblright \ can of course be thought of as
\textquotedblleft consumption\textquotedblright . Evidently, this is a
rather mechanical way of treating such an activity. An interesting extension
would therefore be to include endogenous consumption decisions made by
(potentially risk averse) individuals with (some) foresight. And perhaps
just as importantly, individual motivation may be much more complex,
involving altruism or spite, inequity aversion and/or morality etc.

To mention but one more potential extension: endogeneity of the process that
creates the opportunities. Arguably, both the values of opportunities and
their arrival rates depend in a positive way on current and past national
wealth. Such endogenous growth by way of positive feed-backs may turn the
present ergodic wealth processes into so-called explosive processes for
which it may be possible to identify stable growth paths. While such
generalizations may raise substantial mathematical challenges, they would be
highly relevant for understanding real-world phenomena and amenable to
numerical computer simulations.

Another avenue for future research is laboratory experiments, much in line
with computer games. One could imagine experiments based on models such as
the ones analyzed here but also more complex versions, with a large number
of human subjects who each is given the role of a specific individual $i$ in
the population.\footnote{%
In fact, it could be interesting to let some individuals in the population
be "robots" using pre-programmed behavior rules.} Just as in our theoretical
framework, individuals would be exogenously and randomly matched for
anonymous strategic interaction --- they would then not be informed of the
identity of the other participants in the interaction at hand, only of the
game strategies and payoffs in question and their own role assignment in
that game (which may be symmetric or asymmetric). At the end of a long such
experimental session, each subject could be paid according to his or her
final wealth.\footnote{%
In order to control for individual risk attitudes, an alternative is to let
wealth be given as lottery tickets, with one big final prize given to the
holder of a uniformly randomly drawn lottery ticket at the end of the
session.} The results from such experiments could then be compared with
theoretical predictions and numerical simulations along the lines given here.

The topics of wealth accumulation and wealth distribution are of course big
in the economics literature. Despite this, the models used in that
literature seem to differ starkly from ours. To the best of our knowledge,
neither the present model framework nor our analytical results are closely
related to other work in the economics literature. Indeed, we would be
grateful for suggestions for references to potentially related work. The
closest branch of economics seems to be the search literature, with perhaps
Rubinstein and Wolinsky (1990) coming nearest us. We hope that
generalizations of the present model framework will enable novel analyses of
the accumulation and distribution of fitness and wealth.

\medskip

\section{Appendix}

\medskip

The proofs in this appendix are adaptations to the present setting of proofs
for more general results in Gibaud (2016).

\subsection{Notation and preliminaries}

Let $N\in \mathbb{N}$, for $N>1$, be population size, and let $F$ denote the
state-space, where either $F=\mathbb{N}_{0}^{N}$ or $F=\mathbb{R}$. Denote
by ${C}_{c}^{\infty }(F)$ the set of real-valued smooth (infinitely
differentiable) functions defined on $F$. Let $C_{b}(F)$ be the space of
continuous and bounded functions from $F$ to $\mathbb{R}$. As usual, $\Vert
.\Vert _{\infty }$ represents the sup norm on bounded functions, and $%
|||.||| $ denotes the operator norm on ${C}_{b}(F)$, that is, $|||L|||\
=\sup_{f\in C_{b}(F)}\frac{\Vert Lf\Vert _{\infty }}{\Vert f\Vert _{\infty }}
$.

A (continuous time) stochastic process $X=\left\langle X(t)\right\rangle
_{t\geq 0}$ in $F$ is a random variable taking values in $D(\mathbb{R}%
_{+},F) $, the space of right-continuous left-limited functions, from $%
[0,+\infty )$ to $F$. We will consider Markov processes with initial
distribution $\nu $, a probability measure on $F$, and transition matrix, to
be called the\textit{\ rate matrix,} $(\mathcal{A}(x,y))_{x,y\in F}$, with $%
\mathcal{A}(x,x)=-\sum\nolimits_{y\in F\backslash \left\{ \ x\right\} }%
\mathcal{A}(x,y) $. For $x\neq y$, $\mathcal{A}(x,y)$ is the transition rate
from state $x$ to state $y$. The generator associated with this matrix is
the bounded linear operator $\mathcal{A}$ that sends all bounded and Borel
measurable functions $f$ from $F$ to $\mathbb{R}$, such that for all $x\in F$%
: 
\begin{equation*}
\mathcal{A}f(x)=\sum\limits_{y\in F}\mathcal{A}(x,y)[f(y)-f(x)].
\end{equation*}%
Conversely, from a given generator one can construct the associated rate
matrix from the factors by which the differences $[f(y)-f(x)]$ are
multiplied in the expression of the generator.

Let thus $\mathcal{A}$ be any rate matrix and $\nu $ any probability measure
on $F$. We will construct a Markov process $X$ as follows. First, let $%
\left\langle Y(n)\right\rangle _{n\in \mathbb{N}}$ be a Markov chain (over
discrete time $n$) in $F$, with no absorbing states, and with initial
distribution $\nu $ and transition matrix $\left( \frac{\mathcal{A}(x,y)}{|%
\mathcal{A}(x,x)|}\right) $. Then $\mathcal{A}(x,x)\neq 0$ since there are
no absorbing states. Let $\Delta _{0}$, $\Delta _{1}$,$\dots $, be
independent and exponentially distributed random variables with mean value
1. These are the time intervals between arrivals, and assume that they are
statistically independent of the chain $Y$. We define the Markov process $%
\left\langle X(t)\right\rangle _{t\geq 0}$ in $F$, with initial distribution 
$\nu $ and generator $\mathcal{A}$, by: 
\begin{equation*}
X(t)=\left\{ 
\begin{array}{ll}
Y(0) & \text{for }0\leq t<\frac{\Delta _{0}}{|\mathcal{A}(Y(0),Y(0))|} \\ 
Y(k) & \text{for }\sum\limits_{j=0}^{k-1}\frac{\Delta _{j}}{|\mathcal{A}%
(Y(j),Y(j))|}\leq t<\sum\limits_{j=0}^{k}\frac{\Delta _{j}}{|\mathcal{A}%
(Y(j),Y(j))|}%
\end{array}%
\right.
\end{equation*}

In the present application, we divide the construction of the Markov process 
$W^{N}=(W_{1}^{N},\dots ,W_{N}^{N})$ in two parts, where Part 1 is the
random matching of pairs of individuals to play the game, and Part 2 the
depreciation of individuals' wealth.

[\textit{Part 1}] At the arrivals of a Poisson process with intensity $%
\lambda =N$ a pair of individuals are drawn from the population to play the
game. At each pairwise match, the two individuals make their choices
simultaneously and independently. If all individuals use the unique ESS
strategy $x^{\ast }$, then they play (D,D) with probability $p_{DD}=\left(
1-v/c\right) ^{2}$, (D,H) or (H,D) with probability $p_{DH}=\left(
1-v/c\right) v/c$, and (H,H) with probability $p_{HH}=\left( v/c\right) ^{2}$%
. The generator associated to this part of the wealth process is $\mathcal{A}%
_{g}^{N}$, with domain ${C}_{b}(\mathbb{N}_{0}^{N})$ and defined for all $%
f\in {C}_{b}(\mathbb{N}_{0}^{N})$ and $w\in \mathbb{N}_{0}^{N}$ by%
\begin{equation}
\mathcal{A}_{g}^{N}\left[ f(w)\right] =N\cdot \sum\limits_{\substack{ %
(i,j)\in \{1,\dots N\}^{2}  \\ i\neq j}}\left( 
\begin{array}{l}
\frac{p_{DD}}{N(N-1)}\cdot \left[ f\left( w+\left( v+\frac{y}{2}\right)
(e_{i}+e_{j})\right) -f(w)\right] \\ 
+\frac{p_{DH}}{N(N-1)}\cdot \left[ f\left( w+\left( v+\frac{y}{2}\right)
e_{i}+\frac{y}{2}e_{j}\right) -f(w)\right] \\ 
+\frac{p_{DH}}{N(N-1)}\cdot \left[ f\left( w+\left( v+\frac{y}{2}\right)
e_{j}+\frac{y}{2}e_{i}\right) -f(w)\right] \\ 
+\frac{p_{HH}}{N(N-1)}\cdot \left[ 
\begin{array}{l}
\frac{1}{2}\left( f\left( w+\left( v+\frac{y}{2}\right) e_{i}+\left( -c+%
\frac{y}{2}\right) e_{j}\right) -f(w)\right) \\ 
+\frac{1}{2}\left( f\left( w+\left( v+\frac{y}{2}\right) e_{j}+\left( -c+%
\frac{y}{2}\right) e_{i}\right) -f(w)\right)%
\end{array}%
\right]%
\end{array}%
\right) ,  \label{Ag}
\end{equation}%
where $\left\{ e_{1},\dots e_{N}\right\} $ is the canonical basis of $%
\mathbb{R}^{N}$. The game interaction between any given two individuals $%
i\neq j$ occurs at the arrivals of a Poisson Process with intensity $%
1/\left( N-1\right) $, a process we denote $\mathcal{N}^{i,j}$.

[\textit{Part 2}] In the population at large, individual wealth depreciation
occurs at the arrivals of an independent Poisson process, also with
intensity $\lambda =N$. At each arrival time, one\ individual $i$\ is
randomly drawn for wealth depreciation\ and looses all or part of his or her
current wealth, $w_{i}$, according to the distribution of the random
variable $Z_{w_{i}}$. The depreciation action has a generator $A_{d}^{N}$\
with domain ${C}_{b}(\mathbb{N}_{0}^{N})$, defined for all $f\in {C}_{b}(%
\mathbb{N}_{0}^{N})$ and $w=(w_{1},\dots ,w_{N})\in \mathbb{N}_{0}^{N}$.
Write $z_{i}(k)$\ for $\mathbb{P}(Z_{w_{i}}=k)$, for $k=0,1,...,w_{i}$. We
have for all $w=(w_{0},\dots ,w_{N})\in \mathbb{N}_{0}^{N}$ and $i=1,...N$
that $\sum_{k=0}^{w_{i}}kz_{i}(k)=-\delta w_{i}$ and $z_{i}(w_{i})\geq
\varepsilon >0$. The depreciation infinitesimal generator is 
\begin{equation}
\mathcal{A}_{d}^{N}[f(w)]=\sum_{i=1}^{N}%
\sum_{k=0}^{w_{i}}z_{i}(k)[f(w-ke_{i})-f(w)]  \label{ANd}
\end{equation}%
For any given individual $i\in \{1,\dots ,N\}$, the depreciation Poisson
process $\bar{\mathcal{N}}^{i}$ has intensity $1$, and at each arrival of
this process, individual $i$ looses wealth according to the distribution of $%
Z_{i}$. On average, the individual looses $\delta $ times her current
wealth. For any individuals $i$, $k$ and $h$, where $k\neq h$, the processes 
$\bar{\mathcal{N}}^{i}$ and $\mathcal{N}^{k,h}$ are independent.

Hence, the infinitesimal generator of $\left( W^{N}(t)\right) _{t\geq 0}$
is: 
\begin{equation}
\mathcal{A}^{N}\left[ f(w)\right] =\mathcal{A}_{g}^{N}\left[ f(w)\right] +%
\mathcal{A}_{d}^{N}f(w)  \label{AN}
\end{equation}

The following notation will be convenient in the sequel: for all $w\in 
\mathbb{N}_{0}$, write $\Delta f(w)$ for%
\begin{align*}
\Delta f(w)=& \ 2p_{DD}\left[ f\left( w+\frac{v}{2}+\frac{y}{2}\right) -f(w)%
\right] \\
& +p_{DH}\left[ f\left( w+v+\frac{y}{2}\right) -f(w)\right] \\
& +p_{DH}\left[ f\left( w+\frac{y}{2}\right) -f(w)\right] \\
& +p_{HH}\left( \left[ f\left( w+v+\frac{y}{2}\right) -f(w)\right] +\left[
f\left( w-c+\frac{y}{2}\right) -f(w)\right] \right)
\end{align*}

Let $f\in C_{b}(\mathbb{N}_{0})$. Then $f^{\prime }=(f,0,\dots ,0)\in C_{b}(%
\mathbb{N}_{0}^{N})$, and write 
\begin{equation}
L\left( f(w)\right) =\Delta f(w)+\sum_{k=0}^{w}z_{i}(k)\left[ f(w-k)-f(w)%
\right] .  \label{Lfw}
\end{equation}%
Then $L(f(w))=\mathcal{A}f^{\prime }(w)$, the infinitesimal generator of the
Markov process $\left( W_{i}^{N}(t)\right) _{t}$ for each $i\in \{1,\dots
,N\}$.

\subsection{Proof of Proposition \protect\ref{TE}}

Let $N$ be a positive integer, population size, and let $\left\langle
W_{n}^{N}\right\rangle _{n\in \mathbb{N}_{0}}$ be the Markov chain
associated with the Markov process $\left\langle W^{N}\left( t\right)
\right\rangle _{t\in \mathbb{R}_{+}}$. Write $\theta $ for the zero vector $%
\left( 0,\dots ,0\right) \in \mathbb{N}^{N}$. It is sufficient to show that $%
\left\langle W_{n}^{N}\right\rangle _{n\in \mathbb{N}_{0}}$ is irreducible,
aperiodic and positively recurrent.

That it is irreducible follows from two observations, where the first is
that the probability is positive that all wealth disappears from the
population in period $n=N$, irrespective of the initial state: $\mathbb{P}%
_{w}\left[ W^{N}(N)=\theta \right] >0$ for all $w\in \mathbb{N}^{N}$. The
second observation is that $\left\langle W_{n}^{N}\right\rangle _{n\in 
\mathbb{N}_{0}}$ has exactly one equivalence class, namely, the set 
\begin{equation*}
E=\left\{ w\in \mathbb{N}^{N}:\exists n\in \mathbb{N}\ \text{s.t.}\ \mathbb{P%
}_{\theta }\left[ W_{n}^{N}=w\right] >0\right\} .
\end{equation*}%
This establishes irreducibility. Moreover, since $\mathbb{P}_{\theta }\left(
W_{1}^{N}=\theta \right) \geq \varepsilon >0$, the chain is also aperiodic.

To complete the proof of ergodicity, it thus remains to show that $%
\left\langle W_{n}^{N}\right\rangle _{n}$ is positively recurrent. For this
purpose, one can use Lemma 6.3.20 in Bremaud (2020), which states that if $%
\left( X_{n}\right) _{n}$\textit{\ }is an irreducible Markov chain, $F$\ a
finite subset of its state space $E$, and $\tau (F)$\ is the return time to $%
F$, then then the chain is positive recurrent if $\mathbb{E}_{j}\left[ \tau
(F)\right] )<\infty $\ for all states $j\notin F$. We use this lemma for $E=%
\mathbb{N}_{0}^{N}$ and $F=\{(0,\dots ,0)\}$. The probability that the
wealth of $i$ is depreciated is $\frac{\varepsilon }{2N}$ at each arrival of
the Poisson process. Let $D_{i}$ be the event that the wealth of individual $%
i$ depreciates. Then, for all $\sigma \in \mathbb{N}_{0}^{N}$: 
\begin{equation*}
\mathbb{P}_{\sigma }(\tau \leq N)\geq \mathbb{P}_{\sigma }(\tau =N)\geq 
\mathbb{P}_{\sigma }(D_{1},D_{2},\dots ,D_{N})\geq \left( \frac{\varepsilon 
}{2N}\right) ^{N}.
\end{equation*}%
Thus 
\begin{equation*}
\mathbb{P}_{\sigma }(\tau >N)\leq 1-\left( \frac{\varepsilon }{2N}\right)
^{N}.
\end{equation*}

It follow that for all $\sigma \in \mathbb{N}_{0}^{N}$: 
\begin{align*}
\mathbb{E}_{\sigma }(\tau )& =\sum_{k\in \mathbb{N}}\sum_{l=1}^{N}(Nk+\ell )%
\mathbb{P}_{\sigma }(\tau =kN+\ell ) \\
& \leq \sum_{k\in \mathbb{N}}\sum_{l=1}^{N}(Nk+N)\mathbb{P}_{\sigma }(\tau
>kN) \\
& \leq \sum_{k\in \mathbb{N}}N^{2}(k+1)\mathbb{P}_{\sigma }(\tau >kN)
\end{align*}

But we also have that, for all $\sigma \in \mathbb{N}_{0}^{N}$, 
\begin{align*}
\mathbb{P}_{\sigma }(\tau >kN)& =\mathbb{P}_{\sigma }(\tau >N)\mathbb{P}%
_{\sigma }(\tau >kN|\tau >N) \\
& \leq \left( 1-\left( \frac{\varepsilon }{2N}\right) ^{N}\right)
\sum_{\sigma _{1}\in \mathbb{N}_{0}^{N}}\mathbb{P}_{\sigma }(\tau
>kN,W_{N}^{N}=\sigma _{1}|\tau >N) \\
& \leq \left( 1-\left( \frac{\varepsilon }{2N}\right) ^{N}\right)
\sum_{\sigma _{1}\in \mathbb{N}_{0}^{N}}\mathbb{P}_{\sigma }(\tau
>kN|W_{N}^{N}=\sigma _{1},\tau >N)\mathbb{P}_{\sigma }(W_{N}^{N}=\sigma
_{1}|\tau >N)
\end{align*}%
By the strong Markov property we obtain: 
\begin{align*}
\mathbb{P}_{\sigma }(\tau >kN)& \leq \left( 1-\left( \frac{\varepsilon }{2N}%
\right) ^{N}\right) \sum_{\sigma _{1}\in \mathbb{N}_{0}^{N}}\mathbb{P}%
_{\sigma _{1}}(\tau >(k-1)N)\mathbb{P}_{\sigma }(W_{N}=\sigma _{1}|\tau >N)
\\
& \left( 1-\left( \frac{\varepsilon }{2N}\right) ^{N}\right) ^{k}\prod_{\ell
=1}^{k}\underbrace{\sum_{\sigma _{\ell }\in \mathbb{N}_{0}^{N}}\mathbb{P}%
_{\sigma _{\ell -1}}\left( W_{N}^{N}=\sigma _{\ell }|\tau >(\ell -1)N\right) 
}_{=1} \\
& \leq \left( 1-\left( \frac{\varepsilon }{2N}\right) ^{N}\right) ^{k}
\end{align*}

Hence, for all $\sigma \in \mathbb{N}_{0}^{N}$, 
\begin{equation*}
\mathbb{E}_{\sigma }(\tau )\leq \sum_{k\in \mathbb{N}}N^{2}(k+1)\left(
1-\left( \frac{\varepsilon }{2N}\right) ^{N}\right) ^{k}\leq \frac{N^{2}}{%
\left( \frac{\varepsilon }{2N}\right) ^{2N}}<+\infty ,
\end{equation*}%
which establishes that $\left\langle W_{n}^{N}\right\rangle _{n}$ is
positively recurrent.

Thus $\left\langle W_{n}^{N}\right\rangle _{n}$ has a unique invariant
distribution, and it converges to this from all initial distributions (see
e.g. Thm VIII.6.8 in Barbe and Ledoux, 2007).\medskip

\subsection{Proof of Proposition \protect\ref{T2}}

{Let} $\left\langle u_{N}\right\rangle _{N\in \mathbb{N}}$ {be a sequence of
symmetric probability measures} $u_{N}$ on $\mathbb{N}_{0}^{N}$. {Following
Sznitman (1991), we }say that $\left\langle u_{N}\right\rangle _{N\in 
\mathbb{N}}$ is $u$\emph{-chaotic}, with $u$ a probability measure on $%
\mathbb{N}_{0}$, {if, for any finite collection} $\left\{ \phi _{1},\dots
,\phi _{k}\right\} $ {of continuous and bounded functions on} $\mathbb{N}%
_{0} $, 
\begin{equation*}
\lim\limits_{N\rightarrow +\infty }\int\limits_{\mathbb{N}_{0}^{N}}\phi
_{1}(x_{1})\dots \phi _{k}(x_{k})u_{N}\left( dx_{1}\dots dx_{N}\right)
=\prod\limits_{i=1}^{k}\int\limits_{\mathbb{N}_{0}}\phi _{i}(x)u\left(
dx\right) .
\end{equation*}

The meaning of this definition, if {we apply it to }a fixed and finite
number of individuals, when the total number $N$ {of individuals in the
population} goes to infinity, these individuals' {wealth levels become
i.i.d. with distribution} $u$. The infinitesimal generator of $%
(W_{1}^{N},\dots ,W_{N}^{N})$ is $\mathcal{A}_{g}^{N}$ defined in (\ref{AN}%
). This generator has the shape of a particle system with particles (or
individuals) playing $2\times 2$-games.

Associated with $\mathcal{N}^{i,j}(t)$ and $\bar{\mathcal{N}}^{i}(t)$, let
the processes $\mathcal{M}^{i,j}$ and $\bar{\mathcal{M}}^{i}$ be defined by 
\begin{equation*}
\left( \mathcal{M}^{i,j}(t)\right) _{t}=\left( \mathcal{N}^{i,j}(t)-\frac{t}{%
N-1}\right) _{t}\quad \text{and\quad }\left( \bar{\mathcal{M}}^{i}(t)\right)
_{t}=\left( \bar{\mathcal{N}}^{i}(t)-t\right) _{t}
\end{equation*}%
for all $t\geq 0$.

The object of interest for the proof is 
\begin{eqnarray*}
M_{i}^{f,N}(t)\ &=&\ \sum_{j=1,j\neq i}^{N}\int_{0}^{t}\Delta f\left(
W_{i}^{N}(s)\right) \text{d}\mathcal{M}^{i,j}(s) \\
&&\ +\int_{0}^{t}\sum_{k=0}^{W_{i}^{N}(s)}z_{i}(k)\left[ f\left(
W_{i}^{N}(s)-k\right) -f\left( W_{i}^{N}(s)\right) \right] \text{d}\bar{%
\mathcal{M}}^{i}(s)
\end{eqnarray*}

\begin{lemma}
\label{LA1}For any population size $N>1$, and any pair of individuals $i\neq
j$, with $i,j\in \left\{ 1,...,N\right\} $:%
\begin{equation}
\mathbb{E}\left[ M_{i}^{f,N}(t)M_{j}^{f,N}(t)\right] \leq \frac{4t\Vert
f\Vert _{\infty }}{N-1}\ \quad \forall t>0  \label{bound}
\end{equation}
\end{lemma}

\textbf{Proof}: By independence: 
\begin{align*}
\forall i& \neq j:\ \left\langle \bar{\mathcal{M}}^{i},\bar{\mathcal{M}}%
^{j}\right\rangle =0 \\
\forall i& \neq j,\forall k:\ \left\langle \mathcal{M}^{i,j},\bar{\mathcal{M}%
}^{k}\right\rangle =0 \\
\forall i& \neq j,\forall k\neq h:\ \left\langle \mathcal{M}^{i,j},\mathcal{M%
}^{k,h}\right\rangle =0
\end{align*}%
By the Product Rule (Kurtz, 2001) and writing $u^{-}$\ for the left limit: 
\begin{equation*}
\mathbb{E}\left[ M_{i}^{f,N}(t)M_{j}^{f,N}(t)\right] =
\end{equation*}%
\begin{eqnarray*}
&=&\mathbb{E}\left[ M_{i}^{f,N}(0)M_{j}^{f,N}(0)\right] +\mathbb{E}\left(
\int_{0}^{t}M_{i}^{f,N}(u^{-})\text{d}M_{j}^{f,N}(u)\right) +0+\mathbb{E}%
\left( \left\langle M_{i}^{f,N},M_{j}^{f,N}\right\rangle _{t}\right) \\
&=&\mathbb{E}\left( \int_{0}^{t}\Delta f\left[ W_{i}^{N}(u)\right] \Delta f%
\left[ W_{j}^{N}(u)\right] \right) \text{d}\left\langle \mathcal{M}^{i,j},%
\mathcal{M}^{i,j}\right\rangle _{u}
\end{eqnarray*}%
The claimed inequality (\ref{bound}) follows from the fact that $%
\left\langle \mathcal{M}^{i,j}\right\rangle =\mathcal{N}^{i,j}$. \textbf{%
Q.E.D.}

\medskip

The next step is to establish tightness:

\begin{lemma}
\label{LA2}$\left( \mathcal{L}(W_{1}^{N})\right) _{N}$ is tight.
\end{lemma}

\textbf{Proof:}\footnote{%
This is an application to the present setting of the proof of Lemma 4.8 in
Gibaud (2016).} We apply Theorems 3.9.1 and 3.9.4 in Ethier and Kurtz
(2009), and thus have to verify their hypotheses. First, let $\left( 
\mathcal{F}_{t}\right) _{t}$ be a filtration such that $\left\langle
W^{N}(t)\right\rangle _{t}$ is $\mathcal{F}_{t}$-adapted. Since $W_{1}^{N}$
is a Markov process, 
\begin{equation*}
f(W_{1}^{N}(t))-f(W_{1}^{N}(0))-\int_{0}^{t}Lf(W_{1}^{N}(u-))du
\end{equation*}%
is a $\left( \mathcal{F}_{t}\right) _{t}$-martingale. (And this of course
applies to any individual $i$.)

Since the jumping rates and amplitudes are uniformly bounded (w.r.t. $N$),
we have satisfied the compact containment condition, which states that there
for each $\varepsilon >0$ and $T>0$ exists a bounded subset $K\subset 
\mathbb{N}_{0}$ such that 
\begin{equation*}
\inf_{N}\ \mathbb{P}\left( W_{1}^{N}(t)\in K\ \forall t\leq T\right) \geq
1-\varepsilon \text{.}
\end{equation*}

Moreover, for any $p\in (0,+\infty )$ and $T>0$: 
\begin{equation*}
\sup_{N}\mathbb{E}\left( \int_{0}^{T}\left( Lf(W_{1}^{N}(u-))\right)
^{p}du\right) \leq T2^{p}\Vert f\Vert _{\infty
}^{p}[2p_{DD}+2p_{DH}+2p_{HH}+\varepsilon ]<+\infty
\end{equation*}%
Thus the hypotheses of Theorems 3.9.1 and 3.9.4 of Ethier and Kurtz (2009),
hold, and tightness of $\left( \mathcal{L}(W_{1}^{N})\right) _{N}$ follows. 
\textbf{Q.E.D.}

\smallskip

It now follows from Proposition 2.2 in Sznitman (1991) that $\left( \mathcal{%
L}(\mu ^{N})\right) _{N}$ is tight in $\mathcal{P}\left( \mathcal{P}(\mathbb{%
N}_{0})\right) $.

\smallskip

We next turn to martingale theory. Let $\Pi ^{\infty }$ be a limit point of $%
\left( \mathcal{L}(\mu ^{N})\right) _{N}$ and let $\mu $ have distribution $%
\Pi ^{\infty }$. We prove that $\mu $ satisfies a martingale problem with
initial distribution $\nu $. More precisely: with $f\in C_{b}$, with $%
\left\langle X(t)\right\rangle _{t}$ being the canonical process on $D(%
\mathbb{R}_{+},\mathbb{N}_{0})$, and with $L$ being defined in (\ref{Lfw}): 
\begin{equation*}
M^{f}(t)=f(X(t))-f(X(0))-\int_{0}^{t}Lf(X(u))du
\end{equation*}%
defines a $\mu $-Martingale, and $\mu (0)=\nu $ $\Pi ^{\infty }$ a.s. 
\newline

In order to substantiate this claim, let $k\in \mathbb{N}$, for $k>1$, and
let $0\leq t_{1}\leq t_{2}\leq \dots \leq t_{k}\leq t<T$. Let $g\in C_{b}(%
\mathbb{R}_{+}^{k},\mathbb{N}_{0})$ and $f\in C_{b}(\mathbb{R}_{+},\mathbb{N}%
_{0})$. Finally, let\footnote{%
With $<\mu ,\phi >=\int_{F}\phi (x)\mu (dx)$ where $F$ is a Polish space, $%
\mu $ a probability measure, and $\phi $ a bounded measurable function from $%
F$ to $\mathbb{R}$.} $\mathcal{G}:\mathcal{P}(\mathbb{N}_{0})\rightarrow 
\mathbb{R}$ be defined by 
\begin{equation*}
\mathcal{G}\left( R\right) =\left\langle R\ ,\ \left(
M^{f}(T)-M^{f}(t)\right) g(X(t_{1}),\dots ,X(t_{k}))\right\rangle .
\end{equation*}%
Using the same argument as in the proof of Theorem 4.5 in Graham and M\'{e}l%
\'{e}ard (1997): for all $0\leq t_{1}<t_{2}<\dots <t_{k}\leq t<T$ outside a
countable space, denoted $D$, $\mathcal{G}$ is $\Pi ^{\infty }$ a.s.
continuous. Let us show that 
\begin{equation*}
\mathcal{G}(\mu )=0\quad \Pi ^{\infty }\ \text{a.s.}
\end{equation*}%
Indeed, if it is true for all $0\leq t_{1}<t_{2}<\dots <t_{k}\leq t<T$
outside of $D$, and for any continuous and bounded function $g:\mathbb{R}%
_{+}^{k}\rightarrow \mathbb{R}$, then the following claim holds by the
Monotone Class Theorem: For all $A\subset \mathcal{T}_{t}$ (with $\left( 
\mathcal{T}_{t}\right) $ the natural filtration of $D(\mathbb{R}_{+},\mathbb{%
N}_{0})$), 
\begin{equation*}
\left\langle \mu ,M^{f}(T)1_{A}\right\rangle =\left\langle \mu
,M^{f}(t)1_{A}\right\rangle .
\end{equation*}%
So $\mu $ satisfies the above martingale problem. It thus remains to show
that $\mathcal{G}(\mu )=0$ holds $\Pi ^{\infty }$ a.s.

For this purpose, note that 
\begin{equation*}
\mathcal{G}\left( \frac{1}{N}\sum_{i=1}^{N}\delta _{W_{i}^{N}}\right) =\frac{%
1}{N}\sum_{i=1}^{N}\left( f\left( W_{i}^{N}(T)\right) -f\left(
W_{i}^{N}(t)\right) -\int_{t}^{T}Lf\left( W_{i}^{N}(u)\right) du\right)
g_{i}^{N}
\end{equation*}%
where $g_{i}^{N}=g\left( W_{i}^{N}(t_{1}),\dots ,W_{i}^{N}(t_{k})\right) $.

We have for all $t\in (0,T)$ that $f\left[ W_{i}^{N}(T)\right] -f\left[
W_{i}^{N}(t)\right] $ is equal to 
\begin{equation*}
\sum_{j=1,j\neq i}^{N}\int_{t}^{T}\Delta f\left( W_{i}^{N}(u)\right) \text{d}%
\mathcal{N}^{i,j}(u)+\int_{t}^{T}\sum_{k=0}^{W_{i}^{N}(u)}z_{i}(k)\left[
f\left( W_{i}^{N}(u)-k\right) -f\left( W_{i}^{N}(u)\right) \right] \text{d}%
\bar{\mathcal{N}}^{i}(u)
\end{equation*}%
which is equal to 
\begin{equation*}
M_{i}^{f,N}(T)-M_{i}^{f,N}(t)+\int_{t}^{T}Lf\left[ W_{i}^{N}(u)\right] \text{%
d}u
\end{equation*}

So 
\begin{equation*}
\mathcal{G}\left( \frac{1}{N}\sum_{i}^{N}\delta _{W_{i}^{N}}\right) =\frac{1%
}{N}\sum_{i=1}^{N}\left( M_{i}^{f,N}(T)-M_{i}^{f,N}(t)\right) g_{i}^{N}
\end{equation*}%
and 
\begin{equation*}
\mathbb{E}\left( \left\vert \mathcal{G}\left( \frac{1}{N}\sum_{i=1}^{N}%
\delta _{W_{i}^{N}}\right) \right\vert \right) =A,
\end{equation*}%
where 
\begin{equation*}
A=\mathbb{E}\left( \left\vert \frac{1}{N}\sum_{i=1}^{N}\left(
M_{i}^{f,N}(T)-M_{i}^{f,N}(t)\right) g_{i}^{N}\right\vert \right) .
\end{equation*}%
Moreover, 
\begin{align*}
A^{2}& \leq \frac{1}{N^{2}}\mathbb{E}\left[ \left(
\sum_{i=1}^{N}(M_{i}^{f,N}(T)-M_{i}^{f,N}(t)g_{i}^{N})\right) ^{2}\right] \\
& \leq \frac{1}{N^{2}}\sum_{i,j\in \{1,\dots ,N\}}\mathbb{E}\left[ \left(
M_{i}^{f,N}(T)-M_{i}^{f,N}(t)\right) \left(
M_{j}^{f,N}(T)-M_{i}^{f,N}(t)\right) g_{i}^{N}g_{j}^{N}\right]
\end{align*}%
By exchangeability of $\left\langle W_{i}^{N}\right\rangle _{i}$ we get 
\begin{equation*}
A^{2}\leq \frac{1}{N^{2}}\Vert g_{1}\Vert _{\infty }^{2}\mathbb{E}\left[
\left( M_{1}^{f,N}(T)-M_{1}^{f,N}(t)\right) ^{2}\right] +B,
\end{equation*}%
where%
\begin{equation*}
B=\frac{N(N-1)}{N^{2}}\mathbb{E}\left[ g_{1}^{N}g_{2}^{N}\left(
M_{1}^{f,N}(T)-M_{1}^{f,N}(t)\right) \left(
M_{2}^{f,N}(T)-M_{2}^{f,N}(t)\right) \right]
\end{equation*}%
Since $g$ is bounded: 
\begin{equation*}
B<\Vert g_{1}\Vert _{\infty }\Vert g_{2}\Vert _{\infty }\left[ 
\begin{array}{c}
\mathbb{E}\left[ M_{1}^{f,N}(T)M_{2}^{f,N}(T)\right] -\mathbb{E}\left[
M_{1}^{f,N}(T)M_{2}^{f,N}(t)\right] \\ 
-\mathbb{E}\left[ M_{1}^{f,N}(t)M_{2}^{f,N}(T)\right] +\mathbb{E}\left[
M_{1}^{f,N}(t)M_{2}^{f,N}(t)\right]%
\end{array}%
\right]
\end{equation*}%
Conditioning by $\mathcal{F}_{t}$ and applying Lemma \ref{LA1}: $\mathbb{E}%
\left[ M_{1}^{f,N}(t)M_{2}^{f,N}(t)\right] \leq 4t\Vert f\Vert _{\infty }%
\frac{1}{N-1}$. We get that $A^{2}\rightarrow 0$ as $N\rightarrow +\infty $.

To conclude the proof we use Fatou's lemma: 
\begin{equation*}
\mathbb{E}\left( \left\vert \mathcal{G}(\mu )\right\vert \right) \leq
\lim_{N\rightarrow +\infty }\mathbb{E}\left( \left\vert \mathcal{G}(\mu
^{N})\right\vert \right) =0,
\end{equation*}%
from which $\mathcal{G}(\mu )=0$. Since the mapping $X\in \mathcal{D}(%
\mathbb{R}_{+},\mathbb{N}_{0})\mapsto X(0)$ is continuous, we have $\mu
(0)=\delta _{\nu }$ $\Pi ^{\infty }$ a.s.

Finally, since $\left\langle W_{1}^{N}(t)\right\rangle _{t}$ is a Markov
jump process in $\mathbb{N}_{0}$ with the Feller property, we may use
Theorem 4.1 in Ethier and Kurz (2009) to obtain uniqueness for the
martingale problem. (The hypothesis of that theorem is satisfied for any
Markov generator with the Feller property.) Hence, $\delta _{\mu }$ is the
unique limit point of $\left\langle \Pi ^{N}\right\rangle _{N}$ as $%
N\rightarrow +\infty $. By Proposition 2.2 of Sniztman (1991), $\left\langle
W^{N}\right\rangle _{N}$ is $\mu $-chaotic.

This $\mu $-chaoticity gives (\ref{conv}). Since for all $t>0,$ $M^{f}(t)$
is a $\mu $-martingale, and by setting$\ f=1_{k}$ for any $k\in \mathbb{N}%
_{0}$, we get for all $t>0$: 
\begin{equation*}
0=\int_{D\left( \mathbb{R}_{+},\mathbb{N}_{0}\right) }\left[ 1_{X\left(
t\right) =k}-1_{X\left( t\right) =0}-\int_{0}^{t}L1_{k}\left( X\left(
u\right) \right) du\right] \mu \left( dX\right)
\end{equation*}%
So with $\left\langle \tilde{W}(t)\right\rangle _{t}$ a process of law $\mu $%
, we have for all $k\in \mathbb{N}_{0}$:%
\begin{equation*}
\mathbb{P}(\tilde{W}(t)=k)-\mathbb{P}(\tilde{W}(t)=0)=\int_{0}^{t}\mathbb{E}%
\left[ L1_{k}\left( \tilde{W}(t)\right) \right] du
\end{equation*}

When $\mathbb{P}\left( Z_{k}\in \left\{ 0,k\right\} \right) =1$, one obtains
that $\mathbb{P}(\tilde{W}(t)=k)-\mathbb{P}(\tilde{W}(0)=k)$ equals%
\begin{equation*}
\begin{array}{l}
2p_{DD}\cdot \left[ \mathbb{P}\left( \tilde{W}(t)+\frac{v+y}{2}\right) -%
\mathbb{P}(\tilde{W}(t)=k)\right] \\ 
+\ p_{DH}\cdot \left[ \mathbb{P}\left( \tilde{W}(t)+v+\frac{y}{2}=k\right) -%
\mathbb{P}(\tilde{W}(t)=k)\right] \\ 
+\ p_{DH}\cdot \left[ \mathbb{P}\left( \tilde{W}(t)+\frac{y}{2}=k\right) -%
\mathbb{P}(\tilde{W}(t)=k)\right] \\ 
+\ p_{HH}\cdot \left[ \mathbb{P}\left( \tilde{W}(t)+v+\frac{y}{2}=k\right) -%
\mathbb{P}(\tilde{W}(t)=k)+\mathbb{P}\left( \tilde{W}(t)-c+\frac{y}{2}%
=k\right) -\mathbb{P}(\tilde{W}(t)=k)\right] \\ 
+\ \delta \cdot \left[ \mathbf{1}_{k=0}-\mathbb{P}(\tilde{W}(t)=k)\right]%
\end{array}%
\end{equation*}%
This results in equations (\ref{eqn Evol Eq Jorgen}).

\subsection{Proof of Corollary \protect\ref{C2}}

Let $u_{w}=\mathbb{P}(\tilde{W}(t)=w)$. Then, for all $w\in \mathbb{N}$: 
\begin{align*}
\partial _{t}u_{w}& =2\frac{(c-v)^{2}}{c^{2}}u_{w-\frac{v}{2}-\frac{v}{2}}+2%
\frac{(c-v)v}{c^{2}}\left( u_{w-v-\frac{y}{2}}+u_{w-\frac{y}{2}}\right) \\
& \quad +\frac{v^{2}}{c^{2}}u_{w-v-\frac{y}{2}}+\frac{v^{2}}{c^{2}}u_{w+c-%
\frac{y}{2}}-2u_{w}-\delta u_{w} \\
& =u_{w-\frac{v}{2}-\frac{y}{2}}\cdot \frac{2(c-v)^{2}}{c^{2}}+u_{w-v-\frac{y%
}{2}}\left[ \frac{2(c-v)v}{c^{2}}+\frac{v^{2}}{c^{2}}\right] \\
& \quad +u_{w+c-\frac{y}{2}}\cdot \frac{v^{2}}{c^{2}}+\frac{2(c-v)v}{c^{2}}%
\cdot u_{w-\frac{y}{2}}+u_{w}\left[ -2-\delta \right]
\end{align*}

Denote by $m_{1}$ and $m_{2}$ the first and second moments of $\tilde{W}(t)$%
: 
\begin{equation*}
m_{1}=\sum_{w\in \mathbb{Z}}w\cdot u_{w}\qquad m_{2}=\sum_{w\in \mathbb{Z}%
}w^{2}\cdot u_{w}.
\end{equation*}%
Then 
\begin{align*}
\partial _{t}m_{1}& =\sum_{w\in \mathbb{Z}\backslash \{0\}}w\cdot \partial
_{t}u_{w}\medskip \\
& =\frac{2(c-v)^{2}}{c^{2}}\sum_{w\in \mathbb{Z}}w\cdot u_{w-\frac{v}{2}-%
\frac{y}{2}}+\left[ \frac{2(c-v)v}{c^{2}}+\frac{v^{2}}{c^{2}}\right]
\sum_{w\in \mathbb{Z}}w\cdot u_{w-v-\frac{y}{2}} \\
& \hspace{10pt}+\frac{v^{2}}{c^{2}}\sum_{w\in \mathbb{Z}}w\cdot u_{w+c-\frac{%
y}{2}}+\frac{2(c-v)v}{c^{2}}\sum_{w\in \mathbb{Z}}w\cdot u_{w-\frac{y}{2}}+%
\left[ -2-\delta \right] \sum_{w\in \mathbb{Z}}w\cdot u_{w}\medskip \\
& =\frac{2(c-v)^{2}}{c^{2}}\cdot (m_{1}+\frac{v}{2}+\frac{y}{2})+\left[ 
\frac{2(c-v)v}{c^{2}}+\frac{v^{2}}{c^{2}}\right] \cdot (m_{1}+v+\frac{y}{2})
\\
& \hspace{10pt}+\frac{v^{2}}{c^{2}}\cdot (m_{1}-c+\frac{y}{2})+\frac{2(c-v)v%
}{c^{2}}\cdot (m_{1}+\frac{y}{2})+\left[ -2-\delta \right] \cdot m_{1}
\end{align*}

Hence, 
\begin{equation*}
\partial _{t}m_{1}=\frac{v(c-v)}{c} + y -\delta m_{1}
\end{equation*}

Thus 
\begin{equation*}
m_{1}=\frac{v(c-v)}{\delta c}+\frac{y}{\delta }+K_{1}e^{-\delta t}
\end{equation*}%
for some $K_{1}\in \mathbb{R}$. This establishes (\ref{m1}).

Likewise: 
\begin{align*}
\partial _{t}m_{2}& =\sum_{w\in \mathbb{N}}w^{2}\cdot \partial _{t}u_{w} \\
& =\frac{2(c-v)^{2}}{c^{2}}\sum_{w\in \mathbb{N}_{0}}\left( w-\frac{v}{2}-%
\frac{y}{2}+\frac{v}{2}+\frac{y}{2}\right) ^{2}\cdot u_{w-\frac{v}{2}-\frac{y%
}{2}}\  \\
& \hspace{10pt}+\ \left[ \frac{2(c-v)v}{c^{2}}+\frac{v^{2}}{c^{2}}\right]
\sum_{w\in \mathbb{N}_{0}}(w-v-\frac{y}{2}+\frac{y}{2}+v)^{2}\cdot u_{w-v} \\
& \hspace{10pt}+\frac{v^{2}}{c^{2}}\sum_{w\in \mathbb{N}_{0}}(w+c-\frac{y}{2}%
+\frac{y}{2}-c)^{2}\cdot u_{w+c-\frac{y}{2}}\  \\
& \hspace{10pt}+\left[ \frac{2(c-v)v}{c^{2}}\right] \sum_{w\in \mathbb{N}%
_{0}}(w-\frac{y}{2}+\frac{y}{2})^{2}u_{w-\frac{y}{2}} \\
& \hspace{10pt}+\ \left( -2-\delta \right) \sum_{w\in \mathbb{N}%
_{0}}(w)^{2}\cdot u_{w-\frac{y}{2}}
\end{align*}%
\begin{eqnarray*}
&=&\frac{2(c-v)^{2}}{c^{2}}\left( m_{2}+2\left( \frac{v}{2}+\frac{y}{2}%
\right) m_{1}+\left( \frac{v}{2}+\frac{y}{2}\right) ^{2}\right) \\
&&\hspace{10pt}+\ \left[ \frac{2(c-v)v}{c^{2}}+\frac{v^{2}}{c^{2}}\right]
\left( m_{2}+2\left( v+\frac{y}{2}\right) m_{1}+\left( v+\frac{y}{2}\right)
^{2}\right) \\
&&\hspace{10pt}+\ \frac{v^{2}}{c^{2}}\left( m_{2}+2\left( -c+\frac{y}{2}%
\right) m_{1}+\left( -c+\frac{y}{2}\right) ^{2}\right) \\
&&\hspace{10pt}+\ \frac{2(c-v)v}{c^{2}}\left( m_{2}+ym_{1}+\frac{y^{2}}{4}%
\right) \ -\ \left( 2+\delta \right) m_{2} \\
&=&-\delta m_{2}+K_{1}^{\prime }e^{-\delta t}+A+\frac{B}{\delta }
\end{eqnarray*}%
for 
\begin{equation*}
A=\frac{v^{2}}{2c^{2}}\left( 3c^{2}+2vc-v^{2}\right) +\frac{y}{2c}\left(
2vc+yc-v^{2}\right)
\end{equation*}%
and 
\begin{equation*}
B=2\frac{v^{2}}{c^{2}}\left( c^{2}-2cv+2v^{2}\right) +\frac{2y}{c}\left(
vc+yc-2v^{2}\right)
\end{equation*}

Hence, 
\begin{equation*}
m_{2}=K_{2}te^{-\delta t}+K_{3}e^{-\delta t}+\frac{A}{\delta }+\frac{B}{%
\delta ^{2}}
\end{equation*}%
for $K_{2},K_{3}\in \mathbb{R}$.

\medskip

\subsection{Proof of Proposition \protect\ref{mean field}}

From Proposition 1.5 in Ethier and Kurtz (2009) we have that for all $\sigma
\in \mathbb{N}_{0}^{N}$ and any continuous function $f$\ in a real Banach
space 
\begin{equation*}
\frac{d}{dt}\left[ \mathbb{E}_{\sigma }\left( f\left[ W^{N}(t)\right]
\right) \right] =\mathbb{E}_{\sigma }\left[ \mathcal{A}^{N}\left( f\left[
W^{N}(t)\right] \right) \right]
\end{equation*}%
for 
\begin{equation*}
\mathcal{A}^{N}\left[ f(w)\right] =\mathcal{A}_{g}^{N}\left[ f(w)\right] +%
\mathcal{A}_{d}^{N}[f(w)],
\end{equation*}%
with $e^{i}$ being the $i^{th}$ unit vector, $\mathcal{A}_{g}^{N}$ defined
in (\ref{Ag}), and $\mathcal{A}_{d}^{N}$ in (\ref{ANd}).

Let us here take $f:\mathbb{R}^{N}\rightarrow \mathbb{R}$ to be defined by $%
f(x_{1},\dots ,x_{N})=\left( x_{1}+\dots +x_{N}\right) /N$. Then all the $%
N\left( N-1\right) $\ terms in the above sum are identical, and we obtain 
\begin{eqnarray*}
\mathcal{A}_{g}^{N}\left[ f(W^{N}(t))\right] &=&p_{DD}\left( v+y\right)
+p_{DH}\left( v+y\right) +p_{HD}\left( v+y\right) +p_{HH}(v-c+y) \\
&=&v\left( 1-\frac{v}{c}\right) +y.
\end{eqnarray*}

Moreover, for this particular function\textbf{\ }$f$:%
\begin{align*}
\mathcal{A}_{d}^{N}\left[ f\left( W^{N}(t)\right) \right] &
=\sum_{i=1}^{N}\sum_{k=0}^{W_{i}^{N}}z_{i}(k)\left[ f(w-ke_{i})-f(w)\right]
\\
& =\sum_{i=1}^{N}\sum_{k=0}^{W_{i}^{N}}-\frac{k}{N}z_{i}(k) \\
& =\sum_{i=1}^{N}-\delta \frac{W_{i}^{N}}{N}=-\delta \bar{W^{N}}(t)
\end{align*}

Hence, for all $\sigma \in \mathbb{N}_{0}^{N}$ such that $\bar{W}%
^{N}(0)=w_{0}$: 
\begin{align*}
\frac{d}{dt}\mathbb{E}_{\sigma }\left( f\left[ W^{N}(t)\right] \right) & =%
\mathbb{E}_{\sigma }\left( \mathcal{A}^{N}\left( f\left[ W^{N}(t)\right]
\right) \right) \\
& =v\left( 1-\frac{v}{c}\right) +y-\delta \mathbb{E}_{\sigma }(\bar{W}%
^{N}(t))
\end{align*}%
and thus (\ref{ODE}) obtains.

\subsection{Proof of Proposition \protect\ref{Prop strength}}

Consider the game $G\left( v,c,p\right) $, and let $x\in \left[ 0,1\right] $
denote the probability with player 1 uses pure strategy H, and let $y\in %
\left[ 0,1\right] $ be the probability with which player 2 uses strategy H.
Then H is a best reply for player 1 iff%
\begin{equation}
\left( 1-y\right) \frac{v}{2}+y\left[ \left( v+c\right) p-c\right] \geq 0
\label{NE}
\end{equation}%
This inequality holds when $y=0$. It also holds when $y=1$ iff $p\geq
c/\left( v+c\right) $. With individual $i$ in player role 1, and $p=f\left(
w_{i},w_{j}\right) $, $p>c/\left( v+c\right) $ amounts to condition (\ref%
{diff1}). Under this strict inequality, H thus strictly dominates D for
individual $i$ in player role 1.

Now suppose that $p<c/\left( v+c\right) $. Then $x=1$, that is, for player 1
to play H, is a best reply iff $y\leq y^{\ast }$, where $y^{\ast }$ is
defined by equality in (\ref{NE}): 
\begin{equation*}
y^{\ast }=\frac{v}{v+2c-2\left( v+c\right) p}=\frac{v}{2\left( v+c\right)
f(w_{j},w_{i})-v},
\end{equation*}%
where the last equality presumes that individual $i$ is in player role 1.
Likewise, $y=1$ is a best reply for player 2 iff $x\leq x^{\ast }$, where%
\begin{equation*}
x^{\ast }=\frac{v}{v+2c-2\left( v+c\right) \left( 1-p\right) }=\frac{v}{%
2\left( v+c\right) f(w_{i},w_{i})-v}.
\end{equation*}%
Hence, there are three Nash equilibria: $\left( H,D\right) $, $\left(
D,H\right) $, and $\left( x^{\ast },y^{\ast }\right) $. This establishes the
claims in the proposition.

\end{document}